\documentclass[onecolumn,showpacs,preprintnumbers,showkeys]{revtex4}
\usepackage{amsthm, amscd, amsfonts, amssymb, graphicx}
\usepackage{dcolumn}
\usepackage{bm}
\usepackage[english]{babel}
\usepackage{footnpag}
\usepackage{nccfoots}
\usepackage[symbol*]{footmisc}

\begin{document}

\title{BCS theory with the external pair potential}
\author{Konstantin V. Grigorishin}
\email{konst.phys@gmail.com} \affiliation{Boholyubov Institute for
Theoretical Physics of the National Academy of Sciences of
Ukraine, 14-b Metrolohichna str. Kiev-03680, Ukraine.}

\begin{abstract}
We consider a hypothetical substance, where interaction between
(within) structural elements of condensed matter (molecules,
nanoparticles, clusters, layers, wires etc.) depends on state of
Cooper pairs: an additional work must be made against this
interaction to break a pair. Such a system can be described by BCS
Hamiltonian with the external pair potential term. In this model
the potential essentially renormalizes the order parameter: if the
pairing lowers energy of the structure the energy gap is slightly
enlarged at zero temperature and asymptotically tends to zero as
temperature rises. Thus the critical temperature of such a
superconductor is equal to infinity formally. For this case the
effective Ginzburg-Landau theory is formulated, where the
coherence length decreases as temperature rises, the GL parameter
and the second critical field are increasing functions of
temperature unlike the standard theory. If the pairing enlarges
energy of the structure then suppression of superconductivity and
the first order phase transition occur.
\end{abstract}



\maketitle

\section{Introduction}\label{intr}

Critical temperature $T_{c}$ and the second critical magnetic
field $H_{c2}$ are some of the key characteristics of a
superconductor. These parameters depend on an effective coupling
constant with some collective excitations
$g=\nu_{F}\lambda\lesssim 1$ ($\nu_{F}$ is a density of states at
Fermi level, $\lambda$ is an interaction constant), on frequency
of the collective excitations $\omega$ and on Ginzburg-Landau (GL)
parameter $\kappa$ ($H_{c2}\propto\kappa$). The larger coupling
constant, the larger the critical temperature. For large $g$ we
have $T_{c} \propto \omega\sqrt{g}$ \cite{mahan,ginz} (or $T_{c}
\propto \omega g$ in BCS theory). Formally the critical
temperature can be made arbitrarily large with increasing of the
electron-phonon coupling constant. However in order to reach room
temperature such values of the coupling constant are necessary
which are not possible in real materials. On the other hand we can
increase the frequency $\omega$ due to nonphonon pairing
mechanisms as proposed in \cite{ginz}. However with increasing of
the frequency the coupling constant decreases as $g\propto
1/\omega$. The second critical magnetic field can be enlarged in a
"dirty" limit, where the mean free path $l$ is much less than a
coherence length $l\ll\xi_{0}$ \cite{schmidt}. In this case the GL
parameter can be $\kappa\gg 1/\sqrt{2}$, but still the critical
field is low near the critical temperature: $H_{c2}(T\rightarrow
T_{c})\rightarrow 0$. Many different types of superconducting
materials with a wide variety of electron pairing mechanisms
exist, however all they have the critical temperature limited by
values $\lesssim 100\texttt{K}$, despite the fact that the highly
exotic mechanisms have been proposed.

\begin{figure}[h]
\includegraphics[width=7.7cm]{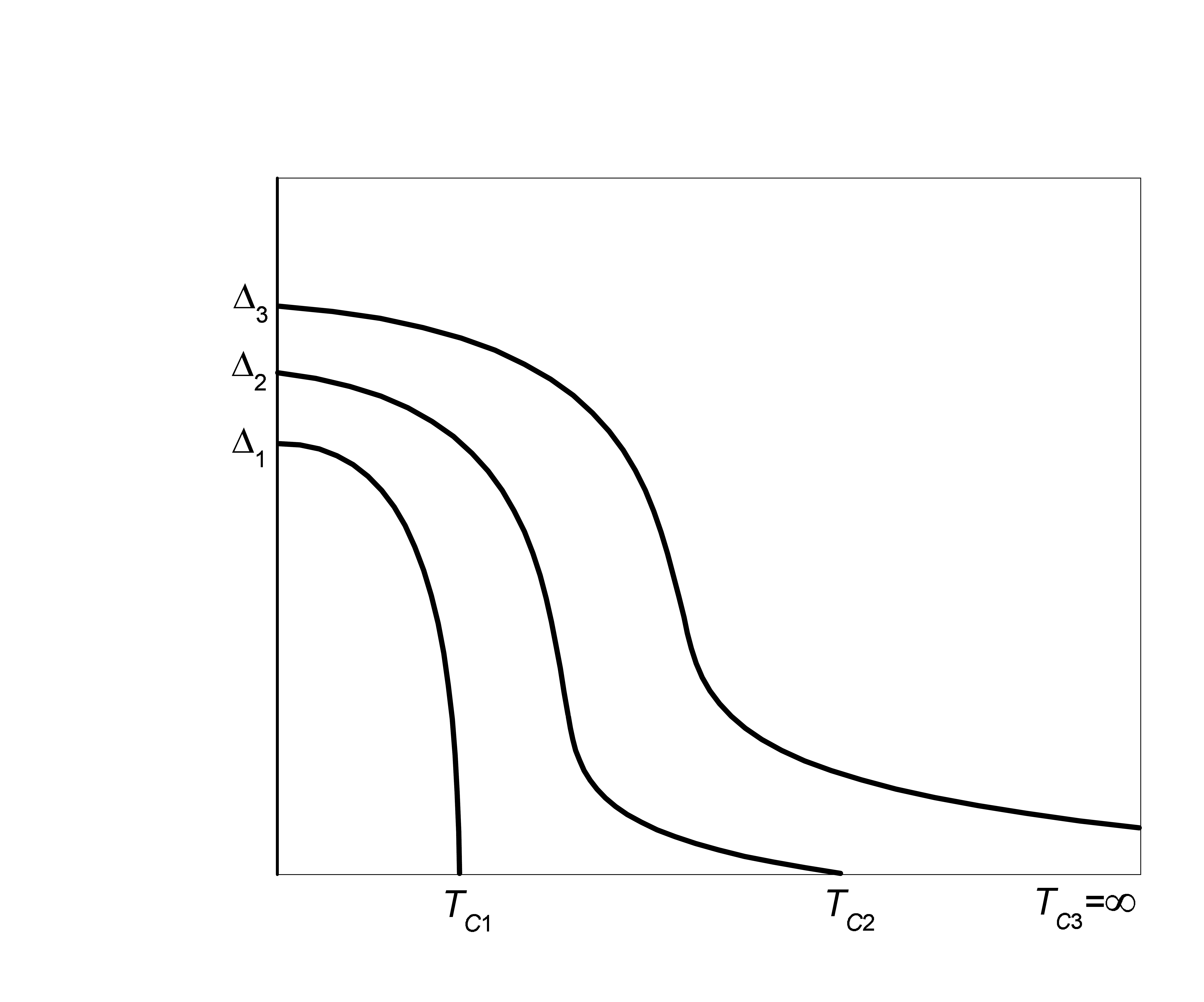}
\caption{Energy gap $\Delta_{1}$ and critical temperature $T_{c1}$
are as in BCS theory: $\frac{2\Delta_{1}}{k_{B}T_{c1}}=3.52$. The
gap $\Delta_{2}$ and the temperature $T_{c2}$ are such that
$\frac{2\Delta_{2}}{k_{B}T_{c2}}<\frac{2\Delta_{1}}{k_{B}T_{c1}}$
as it can be in two-band superconductors \cite{grig,litak}. The
gap $\Delta_{3}(T)$ asymptotically aspires to zero as temperature
rises, so that $\frac{2\Delta_{3}}{k_{B}T_{c3}}=0$. Thus the
critical temperature $T_{c3}$ is equal to infinity in this
hypothetical situation.} \label{Fig1}
\end{figure}

The coupling constant $g$ determines a value of the energy gap at
zero temperature
$\Delta=\hbar\omega/\sinh\left(\frac{1}{g}\right)$, in the same
time a ratio between the gap and the critical temperature is
\begin{equation}\label{0.1}
    \frac{2\Delta}{k_{B}T_{c}}=3.52
\end{equation}
for BCS theory. For presently known materials this ratio is
typically between 3 and 8 \cite{wesche}:  the relatively large
value 8 is considered as evidence for the strong coupling in
high-temperature superconductors, while in conventional metallic
superconductors the ratio is close to 3.5. Thus increasing the
zero temperature gap $\Delta$ we enlarge $T_{c}$. However the
critical temperature can be enlarged by violation the ratio
(\ref{0.1}) as $\Delta/T_{c}\rightarrow 0$ due to some addition
influence on condensate of Cooper pairs as shown in
Fig.(\ref{Fig1}). The simplest examples of such systems are
systems where the proximity effect takes place: two
superconductors are placed in contact \cite{schmidt}, or in
multi-band superconductors where the interband mixing of order
parameters from different bands occurs \cite{asker,grig}. In this
cases the Cooper pairs are injected into the superconductor (band)
from another superconductor (band) with higher critical
temperature (stronger interaction) which plays a role of a source
of Cooper pairs. As a result in a band with lower critical
temperature the ratio (\ref{0.1}) can be violated as
$k_{B}T_{c}>\Delta$ \cite{grig,litak} as illustrated in
Fig.(\ref{Fig1}). However the source is function of temperature
with own critical temperature. In a case of the contact of two
metal the critical temperature of the source drops, however in a
case of the multi-band superconductor the proximity effect
supports superconducting state in all bands.

The hypothetical system where $\Delta/T_{c}=0$ is first proposed
in \cite{loz}. In this work the pairing of massless Dirac
electrons and holes located on opposite surfaces of thin film of
three-dimensional topological insulator is considered. In such a
system both electron-hole Coulomb attraction, which leads to
pairing, and tunneling (or hybridization) of the electrons between
surfaces, which leads to appearance in Hamiltonian the "sources"
of Cooper pairs, act jointly. At $T\rightarrow\infty$ the gap
gradually tends to some value $\Delta_{T}\neq 0$. Thus the
hybridization leads to the smearing of a phase transition into the
paired state in close analogy with the behavior of a ferromagnetic
in an external magnetic field. This solution means the nonzero
energy gap $\Delta_{T}$ if the electron-hole attraction is
switched off or temperature is infinite, which is not physically.
Apparently the gap $\Delta_{T}$ cannot be order parameter, but it
can mean presence of uncorrelated pairs. In its turn the systems
with "preformed" Cooper pairs have been considered in works
\cite{dom,dzum,cap}. In this regime, the electrons are paired, but
they lack the phase coherence necessary for superconductivity. The
existence of preformed pairs implies the existence of a
characteristic energy scale associated to a pseudogap. In a work
\cite{grig1} a model of hypothetical superconductivity has been
proposed, which demonstrates the principal differences from
results of BCS and GL theory due to presence of the source term in
the model Hamiltonian. Unlike the previous works in this model the
coherent paired state is absent if the electron-electron coupling
is absent $g=0$. This means the electron-electron coupling is the
cause of the transition to superconducting state only but not the
source term. In this theory the source term determines energy of a
Cooper pair in some external field which can be called as the
\emph{external pair potential} by analogy with the terminology in
\cite{matt1,dreiz}. In a case of decreasing of Cooper pair's
energy by the potential the energy gap tends to zero
asymptotically as temperature increases. Thus the ratio between
the gap and the critical temperature is $\Delta/T_{c}=0$ instead
of a finite value in BCS theory. In this work the free energy
functional in the limit $T\rightarrow\infty$ has been obtained. It
is shown that the energy gap tends to zero asymptotically as
magnetic field increases. Thus the second critical magnetic field
is equal to infinity as well as the critical temperature, that is
not correct physically, because distance between vortexes cannot
be less than diameter of the vortex's core.

In present work we generalize BCS theory in the sense that the
invariant under $U(1)$ transformation source of Cooper pairs (the
external pair potential) $\widehat{H}_{\upsilon}$ is added to BCS
Hamiltonian. We demonstrate that such Hamiltonian describes a
hypothetical substance, where an interaction energy between
(within) structural elements of condensed matter (molecules,
nanoparticles, clusters, layers, wires etc.) depends on state of
Cooper pairs: an additional work $\upsilon$ must be made against
this interaction to break a pair. In this model the potential
$\upsilon$ essentially renormalizes the order parameter so that
the ratio (\ref{0.1}) changes as $\Delta/T_{c}\rightarrow 0$ if
the breaking of a Cooper pair increases energy of the molecular
structure (or creation of the pair lowers the energy). In another
regime (the breaking of a Cooper pair lowers the energy of the
structure) the renormalization of the order parameter by the
potential $\upsilon$ suppresses a superconducting state and the
first order phase transition occurs. We obtain the free energy of
such a system which generalizes Landau free energy for the
presence of the external pairing potential. For the case
$\Delta/T_{c}\rightarrow 0$ in the limit $T\rightarrow\infty$ we
formulate the effective GL theory which is obtained in
high-temperature limit from the general expression for free energy
(unlike the work \cite{grig1} where the free energy functional was
obtained modifying the ordinary GL expansion). Results of our
effective GL theory radically differ from results of the ordinary
GL theory: the coherence length decreases as temperature rises,
the GL parameter and the second critical field are increasing
functions of temperature.

\section{The model}\label{zero}

\begin{figure}[h]
\includegraphics[width=8.0cm]{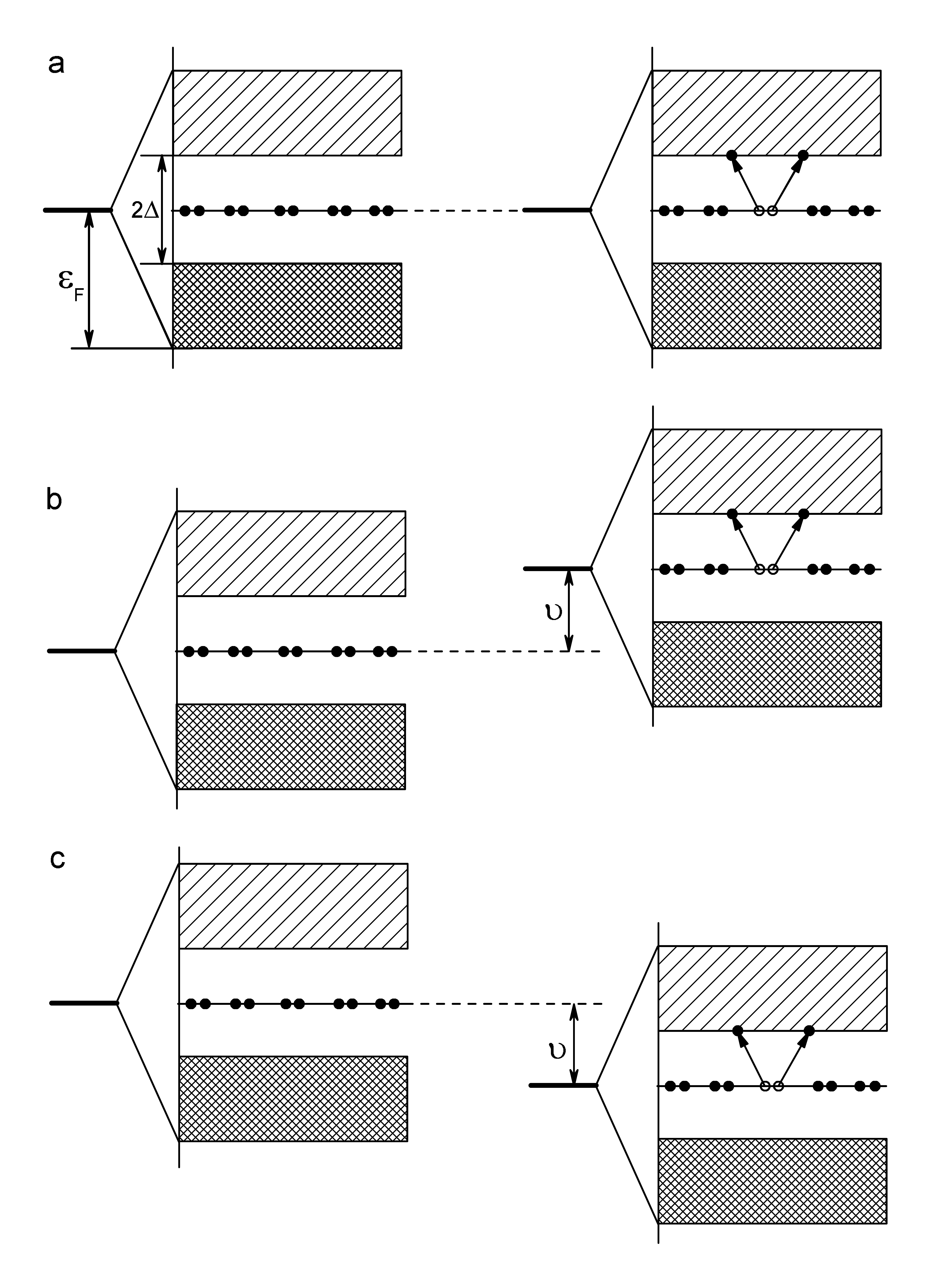}
\caption{ (a) - the process of braking of a Cooper pair in a
conduction band. To break the pair with transfer of its
constituents in the free quasiparticle states the energy
$2|\Delta|$ is needed. (b) - the process of braking of a Cooper
pair if the braking causes the increase of energy of the molecular
structure by $\upsilon$ (the initial energy level shifts). (c) -
it is the same, but if the braking causes the decrease of the
energy.} \label{Fig2}
\end{figure}

According to BCS theory an electron-electron attraction leads to
the appearance of nonzero anomalous averages
$\Delta\sim\left\langle
a_{-\textbf{p}\downarrow}a_{\textbf{p}\uparrow}\right\rangle$ and
$\Delta^{+}\sim\left\langle
a_{\textbf{p}\uparrow}^{+}a_{-\textbf{p}\downarrow}^{+}\right\rangle$,
which are the order parameter (pair potential) of the
superconducting state. The order parameter is determined with some
self-consistency equation $\Delta=I(\Delta)$, which reflects the
fact, that superconductivity is a many-particle cooperative
effect. In this regime the charge is carried by pairs of electrons
(current carriers are the pairs with charge $2e$). To break a pair
with transfer of its constituents in free quasiparticle states the
energy $2|\Delta|$ is needed. We can consider the quantity
$2|\Delta|$ as a work against the effective electron-electron
attraction given the fact that the quantity $\Delta$ is a
collective effect. The superconducting state of a metal and the
breaking of a pair are illustrated in Fig.\ref{Fig2}a.

In our model we consider a hypothetical substance, where an
interaction energy between (within) structural elements of
condensed matter (molecules, nanoparticles, clusters, layers,
wires etc.) depends on state of Cooper pairs: if the pair is
broken, then energy of the molecular system is changed by quantity
$\upsilon=E_{a}-E_{b}$, where $E_{a}$ and $E_{b}$ are energies of
the system after- and before the breaking of the pair accordingly.
Thus to break the Cooper pair we must make the work against the
effective electron-electron attraction and must change the energy
of the structural elements:
\begin{equation}\label{1.1}
    2|\Delta|\longrightarrow 2|\Delta|+\upsilon>0.
\end{equation}
We will call the parameter $\upsilon$ as the \textit{external pair
potential}, since it is imposed on the electron subsystem by the
structural elements of a substance, unlike the pair potential
$\Delta$, which is result of electron-electron interaction and
determined with the self-consistency equation $\Delta=I(\Delta)$.
The parameter $\upsilon$ can be either $\upsilon>0$ or
$\upsilon<0$ and in the simplest case it is not function of the
energy gap $|\Delta|$, $\upsilon=0$ is a trivial case
corresponding to BCS theory. Moreover we suppose that $\upsilon$
does not depend on temperature essentially like parameters of
electron-phonon interaction. The condition $2|\Delta|+\upsilon>0$
ensures stability of the Cooper pairs (bound state of the
electrons is energetically favorable), otherwise transformation
(\ref{1.1}) has no sense and such superconducting state cannot
exist. If $\upsilon<0$ then the breaking of a Cooper pair lowers
energy of the molecular structure (or creation of the pair raises
the energy). In this case the pairs become less stable. If
$\upsilon>0$ then the breaking of the pair increases the energy
(or creation of the pair lowers the energy). In this case the
pairs become more stable. A possible variant of breaking of a
Cooper pair in these cases is illustrated in Fig.\ref{Fig2}b and
Fig.\ref{Fig2}c.

Without going into the details of interaction of the structural
elements we can write an effective Hamiltonian which takes into
account the effect of the structure on Cooper pairs as some
effective external field, like the BCS Hamiltonian is an effective
Hamiltonian describing a system of interacting electrons
independently of nature of this interaction. The order parameter
$\Delta$ is a complex quantity $|\Delta|e^{i\phi}$, where $\phi$
is a phase, and it is the result of a many-particle
self-consistent coherent effect. In the same time the field
$\upsilon$ is an additional parameter imposed on the electron
subsystem by the structural elements. It is easy to see that the
following transformations of the order parameter correspond to the
transformation (\ref{1.1}):
\begin{eqnarray}\label{1.2}
    \Delta\longrightarrow\Delta+\frac{\Delta}{|\Delta|}\frac{\upsilon}{2}=\Delta\left(1+\frac{\upsilon}{2|\Delta|}\right),
    \quad\Delta^{+}\longrightarrow\Delta^{+}+\frac{\Delta^{+}}{|\Delta|}\frac{\upsilon}{2}=\Delta^{+}\left(1+\frac{\upsilon}{2|\Delta|}\right).
   \end{eqnarray}
Really, the work to break the pair is
\begin{eqnarray}
2\sqrt{\Delta\Delta^{+}\left(1+\frac{\upsilon}{2|\Delta|}\right)^{2}}=\left|2|\Delta|+\upsilon\right|.\nonumber
\end{eqnarray}
However the transformations (\ref{1.2}) correspond to the
transformation (\ref{1.1}) if $2|\Delta|+\upsilon>0$ only. If
$2|\Delta|+\upsilon<0$ then the work is $-2|\Delta|+|\upsilon|$.
Thus the model based on the transformations (\ref{1.2}) can give
"parasitic" solutions where the condition (\ref{1.1}) is not
satisfied. Such solutions must be omitted.

The Hamiltonian corresponding to the transformations (\ref{1.2})
is
\begin{eqnarray}\label{1.3}
    \widehat{H}&=&\widehat{H}_{\texttt{BCS}}+\widehat{H}_{\upsilon}=\sum_{\textbf{k},\sigma}\varepsilon(k)a_{\textbf{k},\sigma}^{+}a_{\textbf{k},\sigma}
    -\frac{\lambda}{V}\sum_{\textbf{k},\textbf{p}}a_{\textbf{p}\uparrow}^{+}a_{-\textbf{p}\downarrow}^{+}a_{-\textbf{k}\downarrow}a_{\textbf{k}\uparrow}
    -\frac{\upsilon}{2}\sum_{\textbf{k}}\left[\frac{\Delta}{|\Delta|}a_{\textbf{k}\uparrow}^{+}a_{-\textbf{k}\downarrow}^{+}
    +\frac{\Delta^{+}}{|\Delta|}a_{-\textbf{k}\downarrow}a_{\textbf{k}\uparrow}\right]
\end{eqnarray}
where $\widehat{H}_{\texttt{BCS}}$ is BCS Hamiltonian: kinetic
energy + pairing interaction, energy $\varepsilon(k)\approx
v_{F}(|\textbf{k}|-k_{F})$ is measured from Fermi surface. Indeed,
singling out anomalous averages $\left\langle
    a_{\textbf{p}\uparrow}^{+}a_{-\textbf{p}\downarrow}^{+}\right\rangle,\left\langle
    a_{-\textbf{p}\downarrow}a_{\textbf{p}\uparrow}\right\rangle$ and introducing the order parameter
\begin{eqnarray}\label{1.4}
    \Delta^{+}=\frac{\lambda}{V}\sum_{\textbf{p}}\left\langle
    a_{\textbf{p}\uparrow}^{+}a_{-\textbf{p}\downarrow}^{+}\right\rangle,
    \quad
    \Delta=\frac{\lambda}{V}\sum_{\textbf{p}}\left\langle
    a_{-\textbf{p}\downarrow}a_{\textbf{p}\uparrow}\right\rangle,
\end{eqnarray}
we can rewrite the Hamiltonian in a form
\begin{eqnarray}\label{1.3a}
    \widehat{H}=\sum_{\textbf{k},\sigma}\varepsilon(k)a_{\textbf{k},\sigma}^{+}a_{\textbf{k},\sigma}
    +\sum_{\textbf{k}}\left[\Delta^{+}\left(1+\frac{\upsilon}{2|\Delta|}\right)a_{\textbf{k}\uparrow}a_{-\textbf{k}\downarrow}
    +\Delta\left(1+\frac{\upsilon}{2|\Delta|}\right)
    a_{-\textbf{k}\downarrow}^{+}a_{\textbf{k}\uparrow}^{+}\right]+\frac{1}{\lambda}V|\Delta|^{2}.
\end{eqnarray}
The combinations
$a_{\textbf{k}\uparrow}^{+}a_{-\textbf{k}\downarrow}^{+}$ and
$a_{-\textbf{k}\downarrow}a_{\textbf{k}\uparrow}$ are creation and
annihilation operators of a Cooper pair, thus the mean field
(\ref{1.2}) acts on Cooper pairs. The order parameter is a complex
quantity $\Delta=|\Delta|e^{i\phi}$. Due the multipliers
$\frac{\Delta}{|\Delta|}$ and $\frac{\Delta^{+}}{|\Delta|}$ in
$\widehat{H}_{\upsilon}$ the energy does not depend on the phase
$\phi$ ($a\rightarrow ae^{i\phi/2},a^{+}\rightarrow
a^{+}e^{-i\phi/2}\Longrightarrow \Delta\rightarrow\Delta
e^{i\phi},\Delta^{+}\rightarrow \Delta^{+}e^{-i\phi}$). Thus both
$\widehat{H}_{\texttt{BCS}}$ and $\widehat{H}_{\upsilon}$ are
invariant under the $U(1)$ transformation. The term
$\widehat{H}_{\upsilon}$ is similar to "source term" in
\cite{matt1}, where it means the injection of Cooper pairs into
the system. On the other hand, $\widehat{H}_{\upsilon}$ has a form
of an external field acting on a Cooper pairs only, and $\upsilon$
is energy of a Cooper pair in this field. Eigenfunctions of the
operator $\widehat{H}$ may not be eigenfunctions of a particle
number operator
$\sum_{\textbf{k},\sigma}a_{\textbf{k},\sigma}^{+}a_{\textbf{k},\sigma}$,
that reflects the presence of Cooper pairs' condensate.

In our model we make the following assumptions. A change of the
chemical potential at transition to superconducting state can be
neglected, so that $\mu=\varepsilon_{F}$, because in the model we
suppose $|\Delta|\ll\varepsilon_{F}$, that occurs for systems with
week electron-electron attraction and high density of carriers,
unlike the systems with the strong attraction and low particle
density in the BEC regime, so that
$|\Delta|>\varepsilon_{F},\mu<0$, where the change of the chemical
potential plays important role in formation of the superconducting
state \cite{lok}. In addition we should notice that the coupling
constant $\lambda$ is a sum of phonon term and Coulomb term:
$\lambda=u_{ph}(\omega)-u_{c}$. In metals, as a rule,
$u_{ph}(\omega)<u_{c}$ that corresponds to repulsive
electron-electron interaction, however in such systems the pairing
is possible as result of the second order processes, which lead to
effective attraction regardless of the sign of interaction
\cite{kirz}: $\lambda=u_{ph}(\omega)-u_{c}^{*}$, where
$u_{c}^{*}=u_{c}/\left(1+\nu_{F}u_{c}\ln\frac{\varepsilon_{F}}{\omega}\right)$
is a Coulomb pseudopotential. Such mechanism is not considered in
our model. In the model we suppose a stronger condition
$\lambda=u_{ph}(\omega)-u_{c}>0$, which can occur in nonmetallic
superconductors (for example, in alkali-doped fullerides
$\texttt{A}_{n}\texttt{C}_{60}$, where competition between the
Jahn-Teller coupling and Hund's coupling takes place
\cite{han,nom}).

In order to find equilibrium value of the order parameter $\Delta$
we should obtain system's energy. Using the BCS wave function
$\prod_{\textbf{k}}\left(u(k)+\textsl{v}(k)a_{\textbf{p}\uparrow}^{+}a_{-\textbf{p}\downarrow}^{+}\right)|0\rangle$,
where $u^{2}(k)+\textsl{v}^{2}(k)=1$, we obtain the energy in a
form:
\begin{equation}\label{1.5}
W_{s}=\langle\widehat{H}\rangle=\sum_{\textbf{k}}2\varepsilon(k)\textsl{v}^{2}(k)-\frac{\lambda}{V}\sum_{\textbf{k},\textbf{p}}
u(k)\textsl{v}(k)u(p)\textsl{v}(p)-\upsilon\frac{\Delta}{|\Delta|}\sum_{\textbf{k}}u(k)\textsl{v}(k),
\end{equation}
where $\Delta$ can be supposed real in the absence of magnetic
field and current. The first term is internal kinetic energy, the
second term is energy of the electron-electron interaction, the
third term is energy of Cooper pairs' condensate in the external
pair field. The functions $\textsl{v}(k)$ and $u(k)$ have to
minimize the energy, that is $\frac{dW}{d\textsl{v}^{2}(k)}=0$:
\begin{equation}\label{1.5a}
2\varepsilon(k)-\frac{1-2\textsl{v}^{2}(k)}{u(k)\textsl{v}(k)}\left[\frac{\lambda}{V}\sum_{\textbf{p}}
u(p)\textsl{v}(p)+\frac{\upsilon}{2}\frac{\Delta}{|\Delta|}\right]=0.
\end{equation}
There are two different solutions of this equation, which
correspond two different physical situations:
\begin{enumerate}
    \item We can suppose
\begin{equation}\label{1.6}
\Delta=\frac{\lambda}{V}\sum_{\textbf{p}}u(p)\textsl{v}(p)+\frac{\upsilon}{2}\frac{\Delta}{|\Delta|}.
\end{equation}
Then we obtain the $u,\textsl{v}$ functions in a form
\begin{equation}\label{1.6a}
\textsl{v}^{2}(k)=\frac{1}{2}\left(1-\frac{\varepsilon}{E}\right)\quad
u^{2}(k)=\frac{1}{2}\left(1+\frac{\varepsilon}{E}\right)
\end{equation}
as in BCS theory with
\begin{equation}\label{1.6b}
E^{2}=\varepsilon^{2}+|\Delta|^{2},\quad
\textsl{v}(k)u(k)=\frac{\Delta}{2E}.
\end{equation}
From Eq.(\ref{1.6}) we can see that if the coupling constant is
$\lambda=0$ then the energy gap is nonzero $|\Delta|=\upsilon/2$
(if $\upsilon>0$) that corresponds to results obtained in works
\cite{loz,cap}. Thus the operator $\widehat{H}_{\upsilon}$
describes the external source of Cooper pairs injecting the pairs
into the system. In this case the operator
$\widehat{H}_{\upsilon}$ can have a noninvariant form
$\upsilon\sum_{\textbf{k}}\left[a_{\textbf{k}\uparrow}^{+}a_{-\textbf{k}\downarrow}^{+}
+a_{-\textbf{k}\downarrow}a_{\textbf{k}\uparrow}\right]$. In a
work \cite{matt1} this operator and its analogues are used to
break down the initial symmetry of a noninteracting system and to
produce the desired structure (superconducting, ferromagnetic,
solid etc.) - quasiaverages Bogolyubov method. In the other case
the operator can have a $U(1)\times U(1)$ symmetrical form
$\sum_{\textbf{k}}\left[\upsilon
a_{\textbf{k}\uparrow}^{+}a_{-\textbf{k}\downarrow}^{+}+\upsilon^{+}
a_{-\textbf{k}\downarrow}a_{\textbf{k}\uparrow}\right]$, where
$\upsilon$ is the order parameter of another superconductor, for
example, the boundary (proximity) effect, when a superconductor is
placed in contact with a normal metal \cite{schmidt} or interband
mixing of two order parameters belonging to different bands in a
multi-band superconductor \cite{asker,grig,litak} occurs.
    \item
On the other hand, in order to obtain the functions (\ref{1.6a})
we can suppose
\begin{equation}\label{1.8a}
\Delta=\frac{\lambda}{V}\sum_{\textbf{p}}u(p)\textsl{v}(p),
\end{equation}
then
\begin{equation}\label{1.7}
E^{2}=\varepsilon^{2}+|\Delta|^{2}\left(1+\frac{\upsilon}{2|\Delta|}\right)^{2},\quad
\textsl{v}(k)u(k)=\frac{\Delta\left(1+\frac{\upsilon}{2|\Delta|}\right)}{2E}.
\end{equation}
The integration domain in the self-consistency equation
(\ref{1.8a}) must be cut off at some characteristic phonon energy
$\hbar\omega$ ($\lambda>0$ if $|\varepsilon(k)|<\hbar\omega$,
outside $\lambda=0$), then we have
\begin{equation}\label{1.8}
\Delta=g\int_{-\hbar\omega}^{\hbar\omega}d\varepsilon
\frac{\Delta\left(1+\frac{\upsilon}{2|\Delta|}\right)}{2\sqrt{\varepsilon^{2}+|\Delta|^{2}\left(1+\frac{\upsilon}{2|\Delta|}\right)^{2}}}.
\end{equation}
We can see from Eq.(\ref{1.8}), that if the coupling constant is
$g\equiv\lambda\nu_{F}=0$ then $\Delta=0$. This means, that
\emph{only electron-electron coupling is the cause of
superconductivity but not the potential} $\upsilon$. In this case
the operator $\widehat{H}_{\upsilon}$ is an external field acting
on the Cooper pairs only, and $\upsilon$ is energy of the pair in
this field. As stated above, the potential $\upsilon$ is called as
\textit{external pair potential} since it is imposed on the
electron subsystem by the structural elements of a substance
(molecular, clusters etc. if their energy depends on state Cooper
pairs), unlike the pair potential $\Delta$, which is result of
electron-electron interaction and determined with the
self-consistency equation. If $\upsilon=0$ we have usual
self-consistency equation $\Delta=I(\Delta)$ in BCS theory. In
presence of the external pair potential $\upsilon\neq 0$ the
self-consistency condition has a form $\Delta=I(\Delta,\upsilon)$
- Eq(\ref{1.8}). Thus the potential $\upsilon$ \emph{renormalizes}
the order parameter $\Delta$.

It should be noticed that if we suppose $\Delta=0$, when
$\upsilon\neq 0,g\neq0$, then an uncertainty
$\frac{\Delta}{|\Delta|}=\frac{0}{0}$ takes place in the
self-consistency equation (\ref{1.8}). That is, it would seem,
$\Delta=0$ is not solution, unlike BCS theory. The absence of this
solution is consequence of the fact that the derivative
$\frac{dW}{d\Delta}$ does not exist at $\Delta=0$ because
$W(\Delta\rightarrow 0)\sim |\Delta|$ and the function $|\Delta|$
is non-differentiable in this point (the derivative has a jump
discontinuity), unlike BCS theory where $W(\Delta\rightarrow
0)\sim |\Delta|^{2}$. However, it should be noticed that the order
parameter is a complex quantity: $\Delta=|\Delta|e^{i\phi}$. If
$|\Delta|=0$ then the phase loses the sense so that $\langle
e^{i\phi(\textbf{r},t)}\rangle=0$. Then we have
$\frac{\Delta}{|\Delta|}=0$, hence $\Delta=0$ becomes a solution
of the self-consistency equation (\ref{1.8}) and it means normal
state of a superconductor.

\end{enumerate}

\begin{figure}[h]
\includegraphics[width=8.0cm]{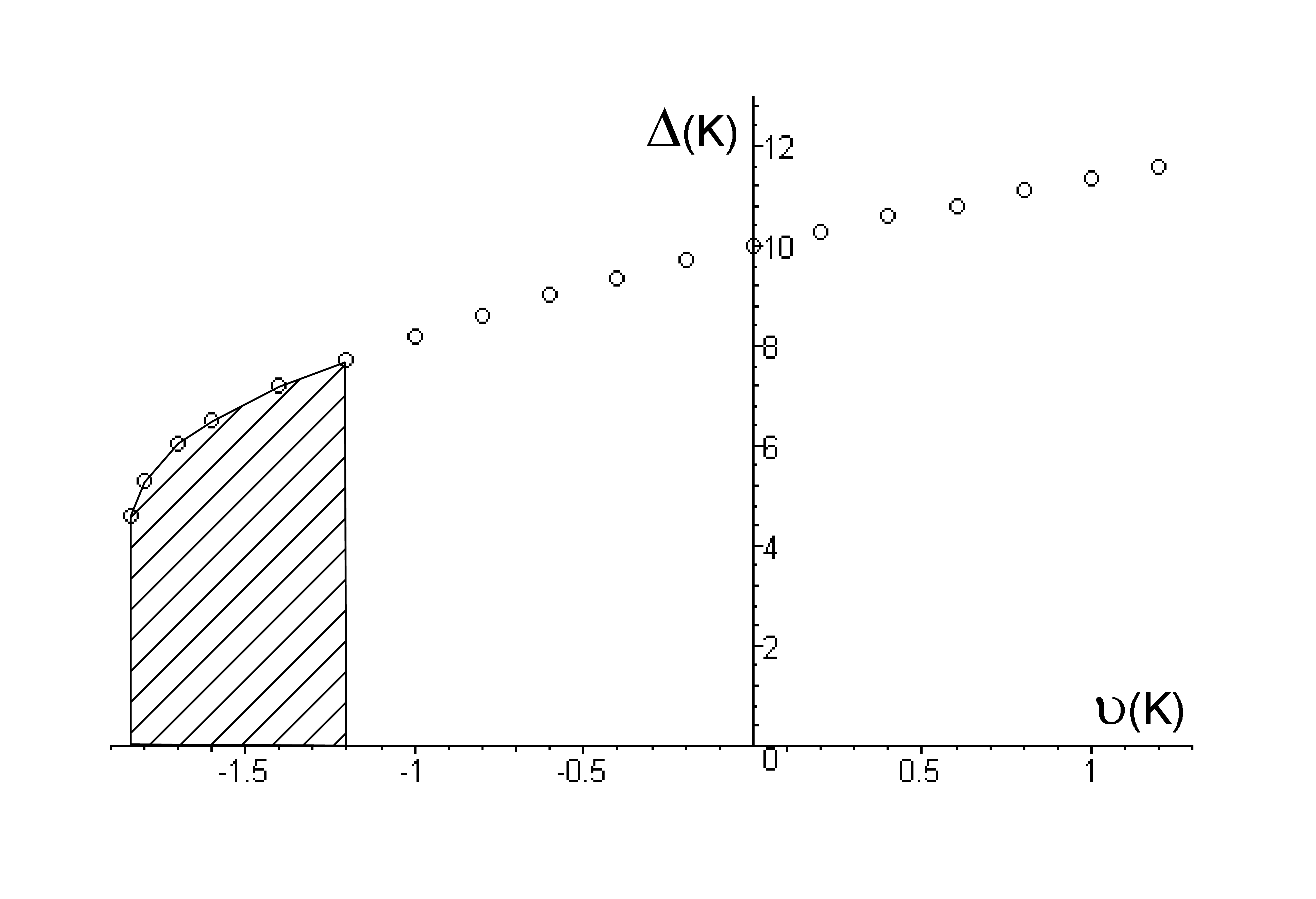}
\caption{the energy gap $|\Delta|$ as function of the potential
$\upsilon$ (in Kelvin) calculated with Eq.(\ref{1.8}), where the
parameters $g=0.25$, $\hbar\omega/k_{B}=273\texttt{K}$ have been
used. If $\upsilon<0$ the critical value
$\upsilon_{c}=-1.84\texttt{K}$ exists, below which the order
parameter is completely suppressed. The shaded region corresponds
to the metastable state, where the energy $W_{s}-W_{n}$ in
Fig.(\ref{Fig5}) is positive.} \label{Fig3}
\end{figure}
\begin{figure}[h]
\includegraphics[width=8cm]{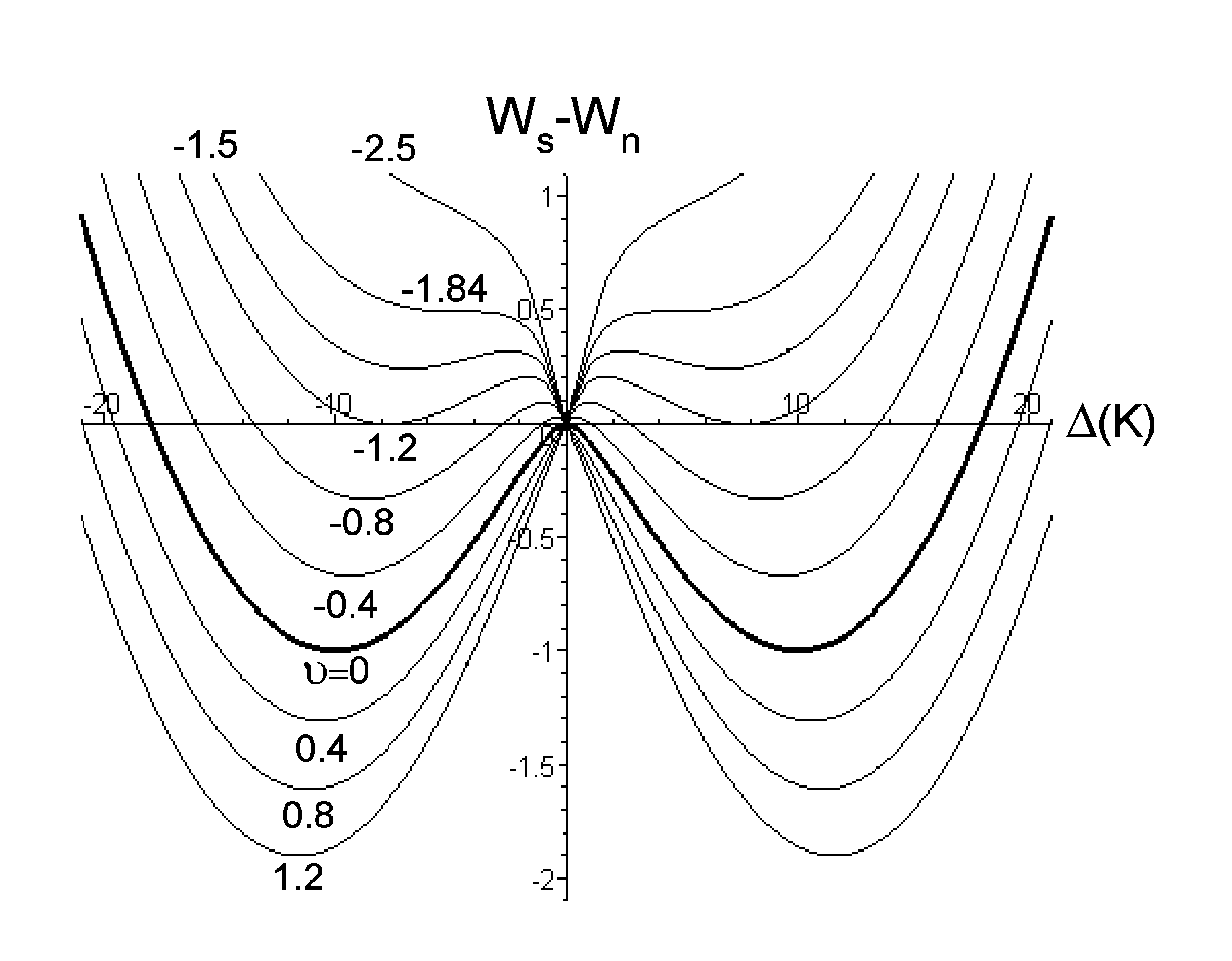}
\caption{the energy (\ref{1.9}) as a function of the energy gap
$\Delta$ at different value of the potential $\upsilon=-2.5,
-1.84,-1.4,-1.2,-0.8,-0.4,0,0.4,0.8,1.2$ in Kelvin. The bold line
is the energy in BCS theory (the external pair potential is absent
$\upsilon=0$). The energy $W_{s}-W_{n}$ is measured by energy of
an equilibrium superconductor without the external pair potential
i.e. $V\nu_{F}\frac{\Delta_{0}^{2}}{2}$, where
$\Delta_{0}=10\texttt{K}$ as it can be seen from Fig.\ref{Fig3}.}
\label{Fig4}
\end{figure}
\begin{figure}[h]
\includegraphics[width=8cm]{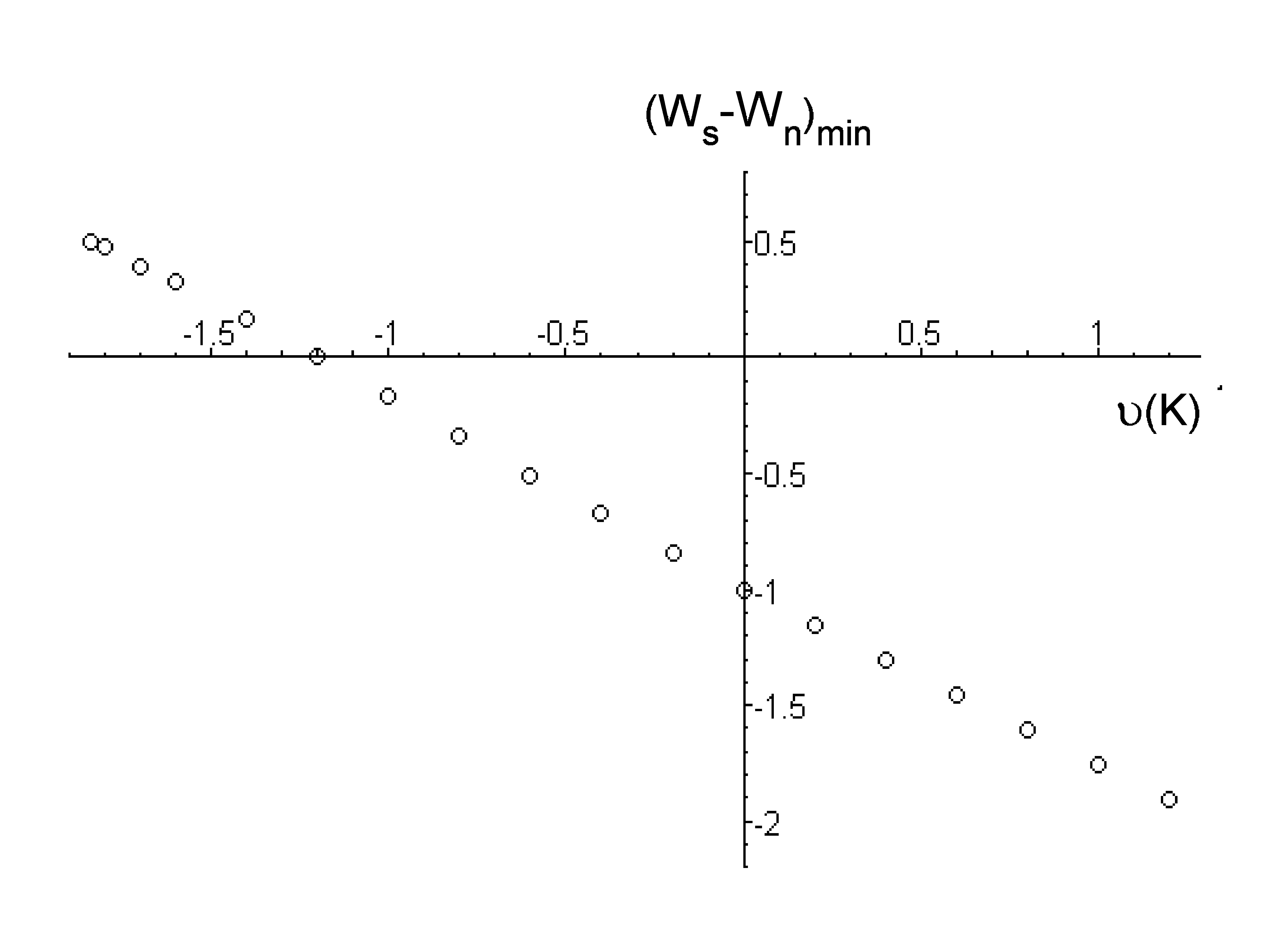}
\caption{The minimums of the energy $W_{s}-W_{n}$ shown in
Fig.\ref{Fig4} corresponding to different values of the potential
$\upsilon$. A region, where $(W_{s}-W_{n})_{min}>0$, corresponds
to metastable states (it is shaded in Fig.\ref{Fig3}).}
\label{Fig5}
\end{figure}

Thus a system described with the Hamiltonian (\ref{1.3}) has the
energy with two minimums (\ref{1.6}) and (\ref{1.8}), which
correspond to two physically different situations: injection of
the Cooper pairs into the system from external source - option 1,
and action of the external pair potential caused by dependence of
energy of the molecular structure on state of a Cooper pair -
option 2. In this work we will consider the model with the
external pair potential only. If $\upsilon>0$ then the pairing
lowers the energy of the molecular structure, that supports
superconductivity $|\Delta(\upsilon>0)|>|\Delta(\upsilon=0)|$. If
$\upsilon<0$ then the pairing of electrons increases the energy of
the molecular structure, that suppresses superconductivity
$|\Delta(\upsilon<0)|<|\Delta(\upsilon=0)|$, moreover such a
critical value of the potential $\upsilon_{c}$ exists, that if
$\upsilon<\upsilon_{c}$ then Eq.(\ref{1.8}) has not any solutions.
As discussed above this region has been supposed as the normal
state of a superconductor. Thus the first order phase transition
(on the parameter $\upsilon$) occurs - Fig.\ref{Fig3}. From the
figure we can see $2|\Delta(\upsilon)|+\upsilon>0$ for all
$\upsilon\geq\upsilon_{c}$, hence the pairs are stable. In the
same time Eq.(\ref{1.8}) has solutions at $g<0,\upsilon<0$,
however it can shown that in this case the condition (\ref{1.1})
is not satisfied, that is the "parasitic" solution occurs in this
case.

Superconducting state is energetically favorable if the energy
(\ref{1.5}) of the superconducting state is less than energy of
normal state i.e. $W_{s}-W_{n}<0$. The energy $W_{n}$ can be
determined as $W_{n}=W_{s}(\Delta=0)$. Using Eq.(\ref{1.5}) we
have:
\begin{eqnarray}\label{1.9}
W_{s}-W_{n}&=&V\nu_{F}\left[\int_{-\hbar\omega}^{\hbar\omega}2\varepsilon
\textsl{v}^{2}(\varepsilon)d\varepsilon-g\left(\int_{-\hbar\omega}^{\hbar\omega}\textsl{v}(\varepsilon)u(\varepsilon)d\varepsilon
\right)^{2}-\upsilon\frac{\Delta}{|\Delta|}\int_{-\hbar\omega}^{\hbar\omega}\textsl{v}(\varepsilon)u(\varepsilon)d\varepsilon\right]\nonumber\\
&-&V\nu_{F}\left[\int_{-\hbar\omega}^{\hbar\omega}2\varepsilon
\textsl{v}_{0}^{2}(\varepsilon)d\varepsilon-g\left(\int_{-\hbar\omega}^{\hbar\omega}\textsl{v}_{0}(\varepsilon)u_{0}(\varepsilon)d\varepsilon
\right)^{2}-\upsilon\int_{-\hbar\omega}^{\hbar\omega}\textsl{v}_{0}(\varepsilon)u_{0}(\varepsilon)d\varepsilon\right],
\end{eqnarray}
where
\begin{equation}\label{1.10}
\textsl{v}^{2}_{0}(\varepsilon)=\frac{1}{2}\left(1-\frac{\varepsilon}{E_{0}}\right),\quad
u^{2}_{0}(\varepsilon)=\frac{1}{2}\left(1+\frac{\varepsilon}{E_{0}}\right),\quad
u_{0}(\varepsilon)\textsl{v}_{0}(\varepsilon)=\frac{\upsilon/2}{2E_{0}},\quad
E_{0}^{2}=\varepsilon^{2}+\left(\frac{\upsilon}{2}\right)^{2}.
\end{equation}
This renormalization of the normal state is a consequence of the
extrapolation in the point $\Delta=0$ where the self-consistent
equation (\ref{1.8}) does not have solutions. The expressions for
the energies $W_{s}$ and $W_{n}$ have not any sense in themselves,
only their difference $W_{s}-W_{n}$ is a physical quantity.
However for a case $\upsilon>0$  interpretation for $W_{n}$ and
$E_{0}$ is possible. Let the electron-electron interaction is
switched off: $g=0$. In addition it can be shown that
$W_{n}<W_{n}(\upsilon=0)$. Then charge is still carried by the
pairs of electrons (current carriers are the pairs) because for
their breaking it must be made the work $\upsilon>0$. This means
that the quasiparticle spectrum has a gap $\upsilon/2$. But this
state is not superconducting because the ordering $\left\langle
aa\right\rangle,\left\langle a^{+}a^{+}\right\rangle$, i.e.
long-range coherence, is absent. In our opinion such state can be
interpreted as state with a pseudogap. However this state
principally differs from the models of pseudogap in high-$T_{c}$
superconductors \cite{dzum,dzum1,dzum2,dzum3,dzum4}, where
preformed uncorrelated pairs are formed due to strong
electron-electron interaction, but the phase coherence is possible
only at more low temperature or at more large concentration of the
pairs. In our model the uncorrelated pairs are pairs in momentum
space (unlike bipolarons which are pairs in real space) which
exist in noninteracting Fermi system due to the external pair
potential. Switching-on of the electron-electron interaction
stipulates the phase coherence. The fermionic nature of the pairs
for such a system is discussed in Appendix \ref{density}. For the
case $\upsilon<0$ such interpretation as "pseudogap" is
impossible, because at $\Delta=0$ the inequality (\ref{1.1}) is
not satisfied.

The energy (\ref{1.9}) is shown in Fig.\ref{Fig4}. If $\upsilon=0$
we have the energy in BCS theory: the energy has symmetric
minimums and a local maximum in a point $\Delta=0$. If
$\upsilon>0$ then the minimums becomes deeper and are located at
larger values of $|\Delta|$. If $\upsilon<0$ then the minimums
becomes shallower and are located at smaller values of $|\Delta|$.
At some value of $\upsilon$ the superconducting state becomes
metastable, because the extremum in $\Delta=0$ becomes a local
minimum with lower energy. At lager value of $|\upsilon|$ the side
minimums disappear and we have only one minimum in $\Delta=0$,
that means the superconducting state is suppressed. The minimums
of the function $W_{s}-W_{n}$ corresponding to different
quantities of the potential $\upsilon$ (that is values of the
function in points determined with Eq.(\ref{1.8})) are shown in
Fig.\ref{Fig5}. We can see the domain of metastable states, where
$(W_{s}-W_{n})_{min}>0$ but Eq.(\ref{1.8}) has solutions. The
region of the solutions $\Delta(\upsilon)$ corresponding to the
metastable states is shaded in Fig.\ref{Fig3}. The minimum of the
energy (\ref{1.9}) can be approximated with the expression:
\begin{eqnarray}\label{1.11}
(W_{s}-W_{n})_{min}=-V\nu_{F}\left[\frac{|\Delta|^{2}}{2}+c\cdot\upsilon|\Delta|\ln\left(\frac{2\hbar\omega}{|\Delta|}\right)\right],
\end{eqnarray}
where $c$ is a coefficient of the order of one and
$|\Delta|\ll\hbar\omega$. If $\upsilon=0$ then we have the energy
as in BCS theory. If $\upsilon<0$ then it can be
$(W_{s}-W_{n})_{min}>0$ at $\Delta\neq 0$, that corresponds to the
metastable states shown in Fig.\ref{Fig3} and Fig.\ref{Fig5}.

\section{Nonzero temperatures}\label{nonzero}

\begin{figure}[h]
\includegraphics[width=8cm]{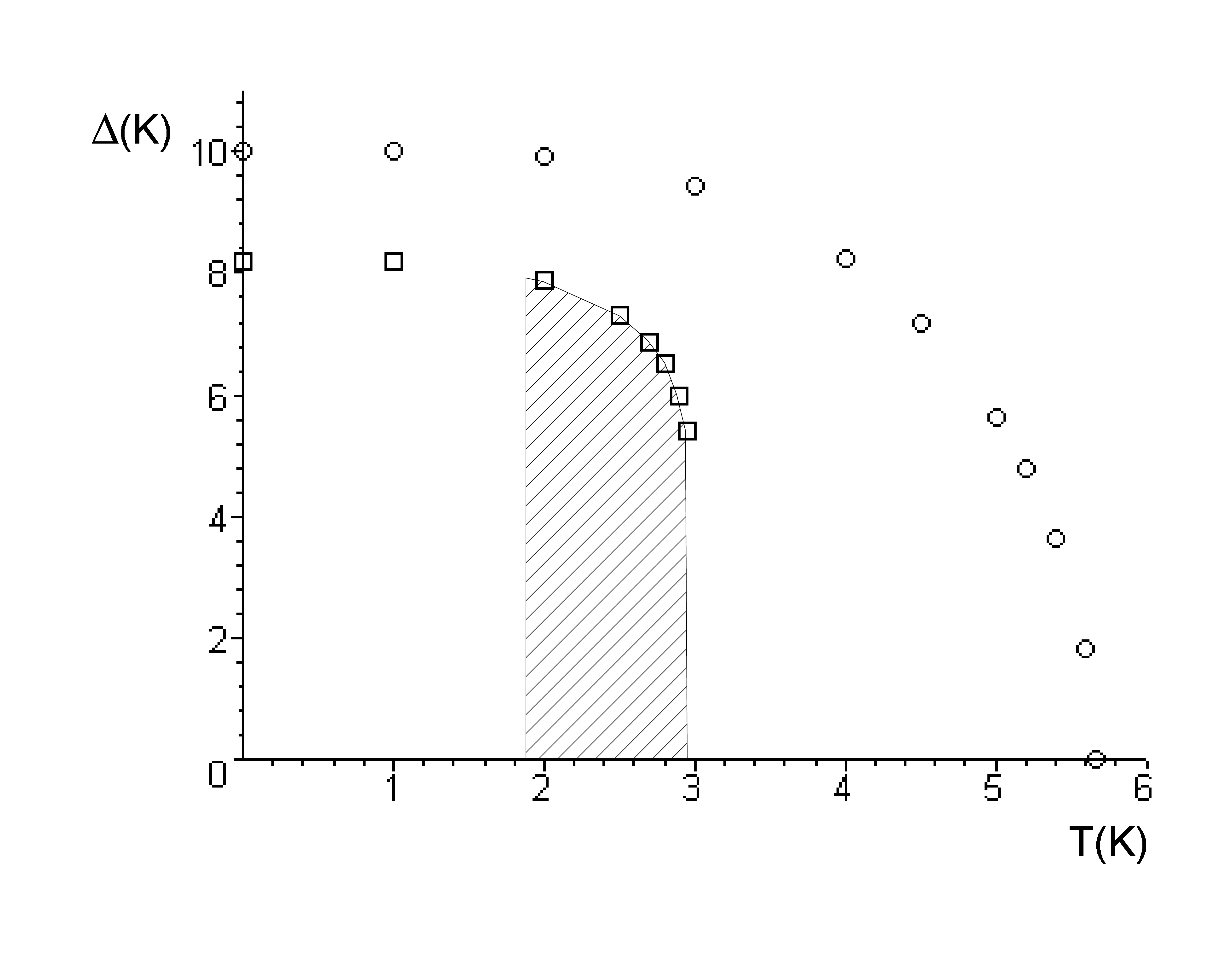}
\caption{the energy gap as a function of temperature (in Kelvin).
Circles are solutions of Eq.(\ref{2.3}) if $\upsilon=0$ - standard
result of BCS theory with the second order phase transition.
Squares are solutions of Eq.(\ref{2.3}) if
$\upsilon=-1\texttt{K}<0$. The solutions demonstrate the first
order phase transition. The shaded region corresponds to the
metastable states, where the free energy $F_{s}-F_{n}$ is
positive.} \label{Fig6}
\end{figure}

At nonzero temperatures thermal activated quasiparticles appears
with some a distribution function $f(k)$, hence the probability
that a pair $(-\textbf{k},\textbf{k})$ can be involved in
formation of superconducting state is $1-2f(k)$ \cite{schmidt}.
Then generalizing Eq.(\ref{1.5}) we obtain the free energy $W-TS$
in a form:
\begin{eqnarray}\label{2.1}
F_{s}&=&\sum_{\textbf{k}}2\varepsilon(k)\left[u^{2}(k)f(k)+\textsl{v}^{2}(k)(1-f(k))\right]-\frac{\lambda}{V}\sum_{\textbf{k},\textbf{p}}
u(k)\textsl{v}(k)u(p)\textsl{v}(p)(1-2f(k))(1-2f(p))\nonumber\\
&+&2k_{B}T\sum_{\textbf{k}}\left[f(k)\ln f(k)+(1-f(k))\ln
(1-f(k))\right]-\upsilon\frac{\Delta}{|\Delta|}\sum_{\textbf{k}}u(k)\textsl{v}(k)(1-2f(k)),
\end{eqnarray}
where the first term is internal kinetic energy, the second term
is energy of the electron-electron interaction, the third term is
contribution of entropy, the last term is energy of Cooper pairs'
condensate in the external pair field. Minimizing the free energy
$\frac{\partial F}{\partial \textsl{v}^{2}(k)}=0$ and
$\frac{\partial F}{\partial f(k)}=0$ we obtain
$\textsl{v}^{2}(k)=\frac{1}{2}\left(1-\varepsilon/E\right)$,
$u^{2}(k)=\frac{1}{2}\left(1+\varepsilon/E\right)$ and
\begin{equation}\label{2.2}
f(k)=\frac{1}{\exp \left(\frac{E}{k_{B}T}\right)+1},\quad
E^{2}=\varepsilon^{2}+|\Delta|^{2}\left(1+\frac{\upsilon}{2|\Delta|}\right)^{2},
\end{equation}
where
\begin{equation}\label{2.3}
\Delta=\frac{\lambda}{V}\sum_{\textbf{p}}u(p)\textsl{v}(p)(1-2f(p))\Longrightarrow
\Delta=g\int_{-\hbar\omega}^{\hbar\omega}d\varepsilon
    \frac{\Delta\left(1+\frac{\upsilon}{2|\Delta|}\right)}{2E}\tanh\left(\frac{E}{2k_{B}T}\right).
\end{equation}
A solutions $\Delta(T)$ of the self-consistency equation
(\ref{2.3}) at $\upsilon<0$ are shown in Fig.\ref{Fig6}, where we
can see suppression of superconductivity
$|\Delta(T,\upsilon<0)|<|\Delta(T,\upsilon=0)|$ and the first
order phase transition. The critical temperature (which is a
function of the parameter $\upsilon$) is less than the critical
temperature of the superconductor without the external pair
potential $T_{c}(\upsilon<0)<T_{c}$. At $T>T_{c}(\upsilon<0)$ the
equation (\ref{2.3}) has only one solution $\Delta=0$ for reasons
discussed in Section \ref{zero}. From Fig.\ref{Fig6} we can see
$2|\Delta(\upsilon)|+\upsilon>0$ for all temperatures $T\leq
T_{c}(\upsilon<0)$, that is the work to break a Cooper pair
(\ref{1.1}) is positive, hence the pairs are stable.

\begin{figure}[h]
\includegraphics[width=8cm]{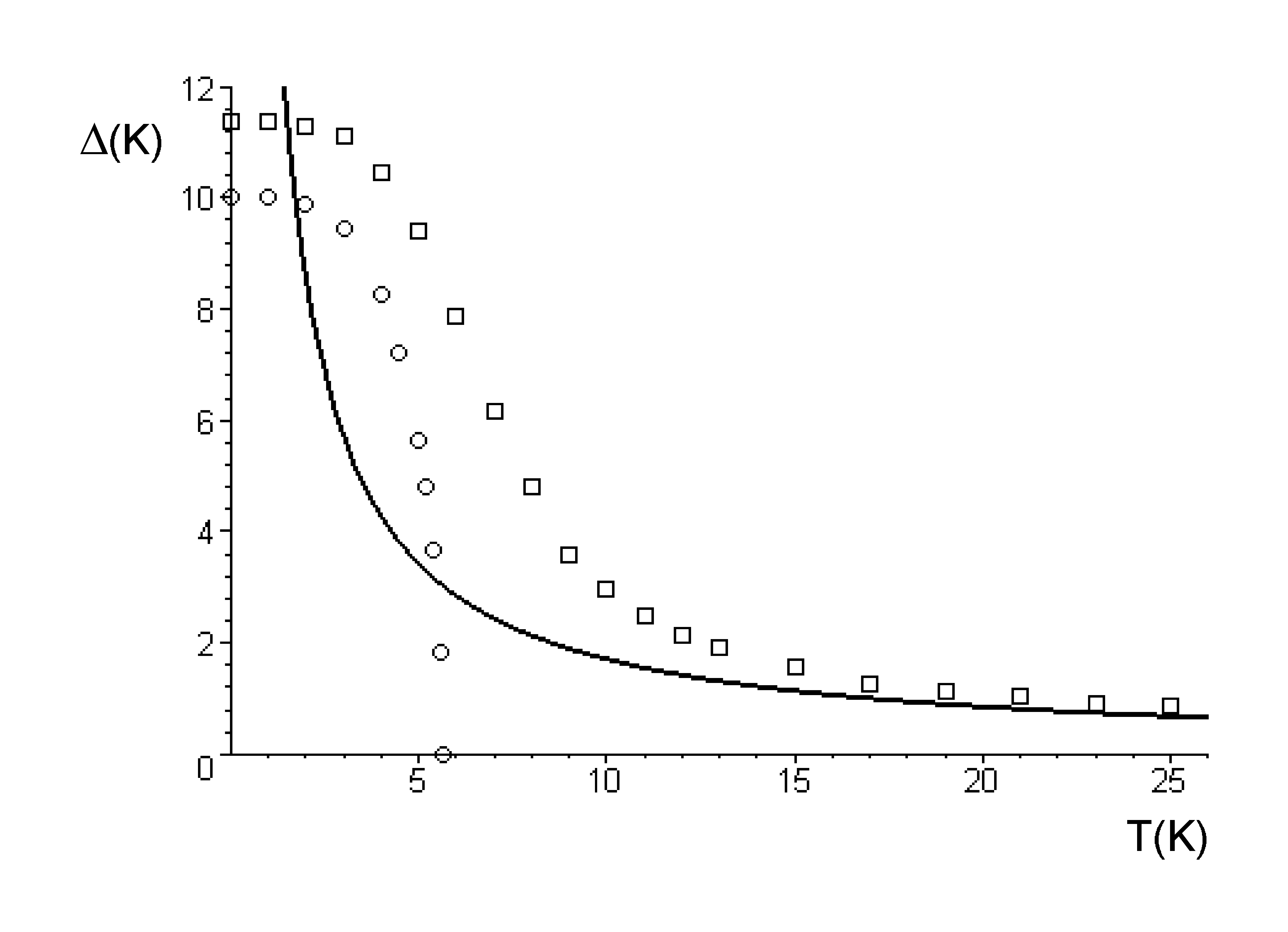}
\caption{the energy gap as a function of temperature (in Kelvin).
Circles are solutions of Eq.(\ref{2.3}) if $\upsilon=0$ - standard
result of BCS theory with the second order phase transition.
Squares are solutions of Eq.(\ref{2.3}) if
$\upsilon=1\texttt{K}>0$. The solutions demonstrate, that the
energy gap tends to zero asymptotically as temperature rises. Bold
line is asymptotical solution (\ref{2.4}) at large temperature.}
\label{Fig7}
\end{figure}

If $\upsilon>0$ then the pairing lowers the energy of the
molecular structure, that supports superconductivity
$|\Delta(T,\upsilon>0)|>|\Delta(T,\upsilon=0)|$. In this case a
solution of Eq.(\ref{2.3}) is such that the order parameter is not
zero at any temperature - Fig.\ref{Fig7}. At large temperature
$T\gg T_{c}$, where $|\Delta(T)|\ll\upsilon$, the gap is
\begin{equation}\label{2.4}
|\Delta|\approx g\int_{-\hbar\omega}^{\hbar\omega}d\varepsilon
    \frac{\upsilon/2}{2\sqrt{\varepsilon^{2}+(\upsilon/2)^{2}}}
    \tanh\left(\frac{\sqrt{\varepsilon^{2}+(\upsilon/2)^{2}}}{2k_{B}T}\right)
    \longrightarrow\frac{g\hbar\omega\upsilon}{4k_{B}T}.
\end{equation}
Formally the critical temperature of such system is equal to
infinity (in reality it limited by the melting of the substance).
It should be noted that if $g=0$, then for any $\upsilon$ a
superconducting state does not exist ($\Delta=0$ always). This
means the electron-electron coupling is the cause of the
transition to superconducting state only but not the external pair
potential.

\begin{figure}[h]
\includegraphics[width=8cm]{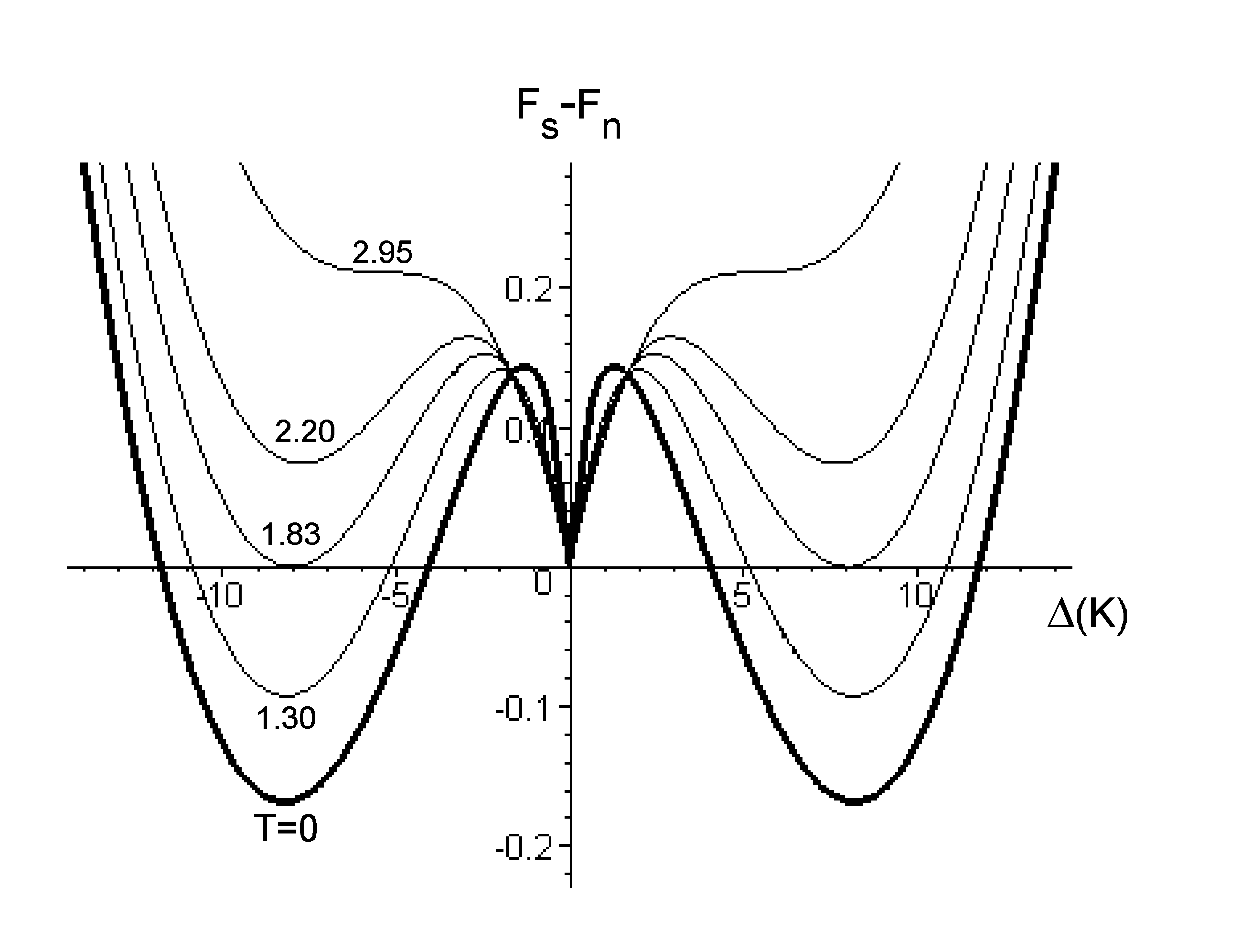}
\caption{the free energy (\ref{2.5}) as a function of the energy
gap $\Delta$ at $\upsilon=-1\texttt{K}$ at different temperatures
$T=0,1.30,1.83,2.20,2.95$ in Kelvin. The energy $F_{s}-F_{n}$ is
measured by energy of an equilibrium superconductor at zero
temperature without the external pair potential i.e.
$V\nu_{F}\frac{\Delta_{0}^{2}}{2}$, where
$\Delta_{0}=10\texttt{K}$ as it can be seen from Fig.\ref{Fig3}.}
\label{Fig8}
\end{figure}

\begin{figure}[h]
\includegraphics[width=15cm]{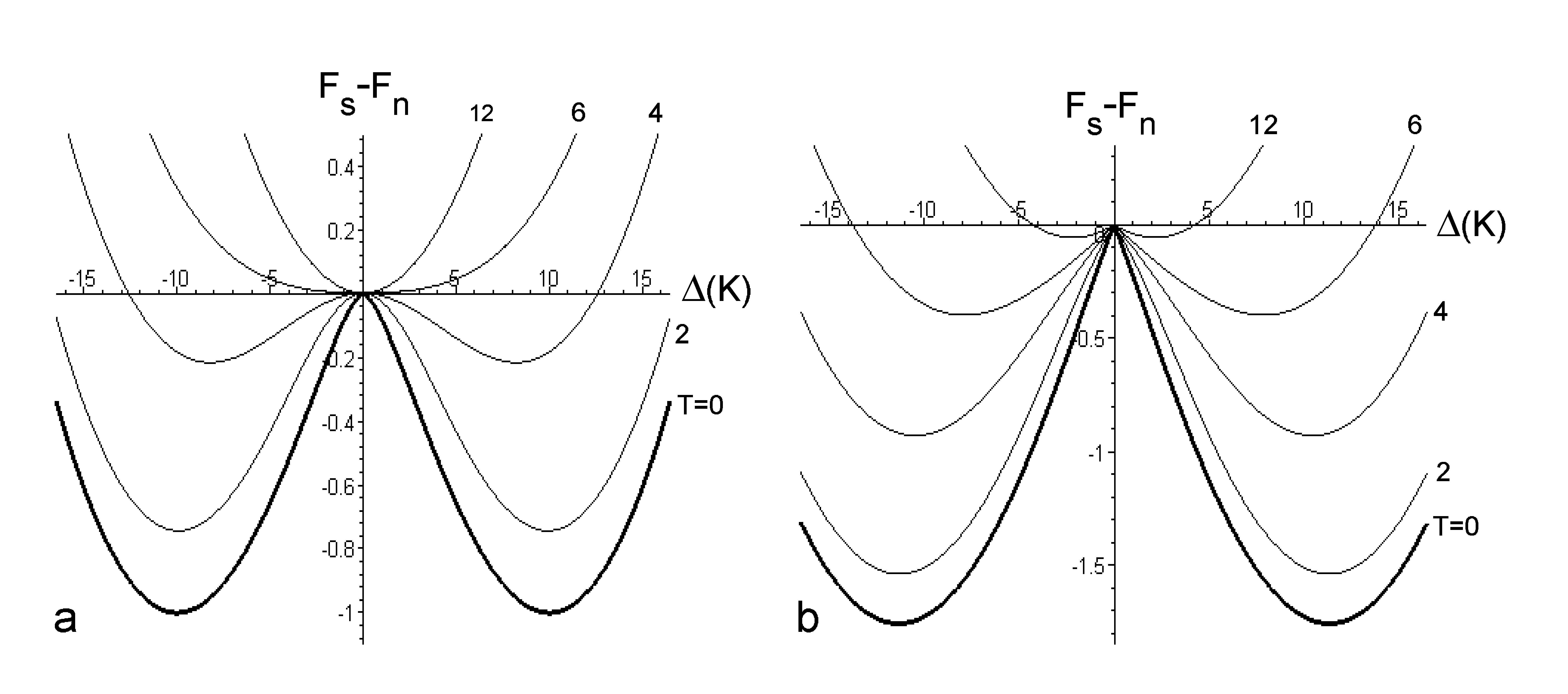}
\caption{(a) - the free energy (\ref{2.5}) as a function of the
energy gap $\Delta$ at absence of the external pair potential
$\upsilon=0$ at different temperatures $T=0,2,4,6,12$ in Kelvin.
(b) - the free energy (\ref{2.5}) at $\upsilon=1\texttt{K}$ at the
same temperatures $T=0,2,4,6,12$. As in Fig.\ref{Fig8} the energy
$F_{s}-F_{n}$ is displayed as dimensionless value.} \label{Fig9}
\end{figure}

Superconducting state is energetically favorable if the free
energy (\ref{2.1}) is less than the free energy of normal state
i.e $F_{s}-F_{n}<0$. The free energy $F_{n}$ can be determined as
$F_{n}=F_{s}(\Delta=0)$. Then using Eq.(\ref{2.1}) and the method
of Section \ref{zero} - Eqs.(\ref{1.9},\ref{1.10}), we can write
the free energy in a form:
\begin{eqnarray}\label{2.5}
F_{s}-F_{n}&=&V\nu_{F}[\int_{-\hbar\omega}^{\hbar\omega}2\varepsilon
\left[u^{2}f+\textsl{v}^{2}(1-f)\right]d\varepsilon-g\left(\int_{-\hbar\omega}^{\hbar\omega}\textsl{v}u(1-2f)d\varepsilon
\right)^{2}\nonumber\\
&+&2k_{B}T\int_{-\hbar\omega}^{\hbar\omega}\left[f\ln f+(1-f)\ln
(1-f)\right]d\varepsilon-\upsilon\frac{\Delta}{|\Delta|}\int_{-\hbar\omega}^{\hbar\omega}\textsl{v}u(1-2f)d\varepsilon]\nonumber\\
&-&V\nu_{F}[\int_{-\hbar\omega}^{\hbar\omega}2\varepsilon
\left[u^{2}_{0}f_{0}+\textsl{v}^{2}_{0}(1-f_{0})\right]d\varepsilon-g\left(\int_{-\hbar\omega}^{\hbar\omega}\textsl{v}_{0}u_{0}(1-2f_{0})d\varepsilon
\right)^{2}\nonumber\\
&+&2k_{B}T\int_{-\hbar\omega}^{\hbar\omega}\left[f_{0}\ln
f_{0}+(1-f_{0})\ln
(1-f_{0})\right]d\varepsilon-\upsilon\int_{-\hbar\omega}^{\hbar\omega}\textsl{v}_{0}u_{0}(1-2f_{0})d\varepsilon],
\end{eqnarray}
where
\begin{equation}\label{2.6}
f_{0}(\varepsilon)=\frac{1}{\exp
\left(\frac{E_{0}}{k_{B}T}\right)+1} ,\quad
E_{0}^{2}=\varepsilon^{2}+(\upsilon/2)^{2}.
\end{equation}
If $\upsilon=0$ we have the free energy in BCS theory: the energy
has symmetric minimums and a local maximum in a point $\Delta=0$ -
Fig.\ref{Fig9}a. At $T>T_{c}$ the free energy has only one minimum
in $\Delta=0$ - superconducting state is absent. The free energy
(\ref{2.5}) at different temperatures and $\upsilon=-1\texttt{K}$
is shown in Fig.\ref{Fig8}. With increasing of temperature the
minimums becomes shallower and are located in smaller values of
$|\Delta|$. At some temperature the superconducting states in the
minimums becomes metastable, because the extremum in $\Delta=0$
becomes a minimum with lower energy. In the metastable phase the
free energy $(F_{s}-F_{n})_{min}>0$ but Eq.(\ref{2.3}) has
solutions. At higher temperature these minimums disappear, and we
have only one minimum in $\Delta=0$, that means the
superconducting state is absent. The region of the solutions
$\Delta(T)$ corresponding to the metastable states is shaded in
Fig.\ref{Fig6}.

If $\upsilon>0$ then the minimums become deeper and they are
located in larger values of $|\Delta|$ than in the case
$\upsilon=0$ - Fig.\ref{Fig9}b. However the main difference
between these cases is that the free energy at $\upsilon>0$ has
absolute minimums where $(F_{s}-F_{n})_{min}<0$ at all
temperatures. This means the system is in superconducting state at
any temperature. The minimums of the free energy (\ref{2.5}) at
$T\rightarrow\infty$ are determined with Eq.(\ref{2.4}).


Let us consider several important limit cases. Let the external
pair potential is absent $\upsilon=0$ and temperature is slightly
less than the critical temperature $T\lesssim T_{c}$, hence the
gap is small $\Delta\rightarrow 0$. Then we can expand the free
energy (\ref{2.5}) in powers of the order parameter:
\begin{equation}\label{2.7}
F_{s}-F_{n}=V\nu_{F}\left[a(T-T_{c})|\Delta|^{2}+\frac{b}{2}|\Delta|^{4}\right],
\end{equation}
where $a$ and $b$ are some coefficient. Thus we have
Ginzburg-Landau expansion.

The case $\upsilon>0$ in a limit $T\rightarrow\infty$ is the most
interesting. The energy gap $\Delta$ is small in this region. Then
we can expand the free energy (\ref{2.5}) in powers of $1/T$ and
in powers of $|\Delta|$:
\begin{equation}\label{2.8}
F_{s}-F_{n}=V\nu_{F}\left[\frac{\hbar\omega}{2k_{B}T}|\Delta|^{2}-\upsilon
g\left(\frac{\hbar\omega}{2k_{B}T}\right)^{2}|\Delta|\right].
\end{equation}
From condition $\frac{dF}{d|\Delta|}=0$ we obtain
$|\Delta|=\frac{g\hbar\omega\upsilon}{4k_{B}T}$ that coincides
with Eq.(\ref{2.4}). It should be noted that in this limit the
contribution of the kinetic energy (the first term in
Eq.(\ref{2.1})) is proportional to $1/T^3$, in the same time the
contribution of the entropy (the third term in Eq.(\ref{2.1})) is
proportional to $1/T$. Thus main factor, which resists the pairing
at $T\rightarrow\infty$, is the entropy, and the contribution of
the kinetic energy can be omitted unlike GL expansion.
Substituting the energy gap (\ref{2.4}) in the function
(\ref{2.8}) we have a gain in the free energy for the
superconducting state:
\begin{equation}\label{2.9}
\left(F_{n}-F_{s}\right)_{min}=V\nu_{F}\frac{g^{2}\upsilon^{2}(\hbar\omega)^{3}}{16(k_{B}T)^{3}}=\frac{\mu_{0}H_{cm}^{2}}{2}V
\Longrightarrow H_{cm}\propto\frac{\upsilon}{T^{3/2}},
\end{equation}
that determines the thermodynamical magnetic field $H_{cm}$, which
tends to zero asymptotically as temperature rises and is
proportional to the potential $\upsilon$.

\section{Free energy functional}\label{free}

At presence of gradient of the order parameter
$\overrightarrow{\nabla}\Delta\neq 0$ (for example, when a current
$\textbf{j}=2en_{s}\textbf{q}/2m$ occurs) electrons are paired
with momentums $\textbf{k}+\frac{\textbf{q}}{2}$ and
$-\textbf{k}+\frac{\textbf{q}}{2}$ accordingly. The nonzero
momentum of the pairs causes an additional kinetic energy of the
condensate, hence the expansion of superconductor's free energy
has a form \cite{schmidt}:
\begin{equation}\label{3.1}
    F_{s}-F_{n}=-\alpha
    n_{s}+\frac{\beta}{2}n_{s}^{2}+n_{s}\frac{q^{2}}{4m},
\end{equation}
where $n_{s}$ is the condensate's density. In this case the
density is a function of the momentum, and a critical momentum
$q_{c}=\sqrt{4m\alpha}$ exists such that $n_{s}(q>q_{c})=0$. As a
rule $q_{c}\propto\frac{m|\Delta|}{k_{F}}\ll k_{F}$. Thus we have
to consider an operator of the kinetic energy in a form
\begin{eqnarray}\label{3.2}
\widehat{H}_{kin}=\sum_{\textbf{k}}\left[\varepsilon\left(\textbf{k}+\frac{\textbf{q}}{2}\right)a_{\textbf{k}
+\frac{\textbf{q}}{2}\uparrow}^{+}a_{\textbf{k}+\frac{\textbf{q}}{2}\uparrow}
+\varepsilon\left(-\textbf{k}+\frac{\textbf{q}}{2}\right)a_{-\textbf{k}
+\frac{\textbf{q}}{2}\downarrow}^{+}a_{-\textbf{k}+\frac{\textbf{q}}{2}\downarrow}\right],
\end{eqnarray}
where
\begin{equation}\label{3.3}
\varepsilon\left(\textbf{k}+\frac{\textbf{q}}{2}\right)\approx\varepsilon\left(k\right)+\frac{\textbf{kq}}{2m},\quad
\varepsilon\left(-\textbf{k}+\frac{\textbf{q}}{2}\right)\approx\varepsilon\left(k\right)-\frac{\textbf{kq}}{2m}.
\end{equation}
Using the free energy (\ref{2.5}) with the kinetic energy
calculated on the operator (\ref{3.2}) and passing to the limit
$T\rightarrow\infty,|\Delta|\ll\upsilon$ at $\upsilon>0$ we can
obtain a free energy:
\begin{equation}\label{3.4}
F_{s}-F_{n}=V\left[-A|\Delta|+\frac{B}{2}|\Delta|^{2}+Cq^{2}|\Delta|
\right],
\end{equation}
where the coefficients are
\begin{equation}\label{3.5}
A=\nu_{F}g\left(\frac{\hbar\omega}{2k_{B}T}\right)^{2}\upsilon,\quad
B=\nu_{F}\frac{\hbar\omega}{k_{B}T},\quad
C=\nu_{F}\frac{\hbar\omega
k_{F}^{2}}{144\left(k_{B}T\right)^{3}m^{2}}\upsilon.
\end{equation}
Then we can obtain an equilibrium value of the gap, value of the
free energy in this point and the critical momentum of the
condensate accordingly:
\begin{eqnarray}
&&\frac{\delta F}{\delta|\Delta|}=0\quad\Longrightarrow\quad
|\Delta|_{min}=\frac{A}{B}\left(1-\frac{C}{A}q^{2}\right)\label{3.6}\\
\nonumber\\
&&(F_{s}-F_{n})_{min}=-\frac{A^{2}}{2B}\left(1-\frac{C}{A}q^{2}\right)^{2}\Longrightarrow
q_{c}^{2}=\frac{A}{C}\label{3.7}
\end{eqnarray}

In the general case the vector $\textbf{q}$ can run through the
whole momentum space (when the pairing potential is spatially
inhomogeneous
$\Delta(\textbf{r})=\sum_{\textbf{q}}\Delta_{\textbf{q}}e^{i\textbf{qr}/\hbar}$),
then we should sum the free energy (\ref{3.4}) over all values of
the momentum $\textbf{q}$. Moreover if the system is in a magnetic
field, then the momentum $\textbf{q}$ must be replaced by
$\textbf{q}-2e\textbf{a}_{\textbf{q}}$, where
$\textbf{a}_{\textbf{q}}$ is a Fourier component of a magnetic
vector potential
$\textbf{A}(\textbf{r})=\sum_{\textbf{q}}\textbf{a}_{\textbf{q}}e^{i\textbf{qr}/\hbar}$,
so that $\textbf{B}=\texttt{curl}\textbf{A}$ is a magnetic
induction vector. Then the free energy functional takes a form
\begin{equation}\label{3.8}
F_{s}-F_{n}=V\sum_{\textbf{q}}\left[-A|\Delta_{\textbf{q}}|+\frac{B}{2}|\Delta_{\textbf{q}}|^{2}
+C\left(\textbf{q}-2e\textbf{a}_{\textbf{q}}\right)^{2}|\Delta_{\textbf{q}}|
+\frac{1}{2\mu_{0}\hbar^{2}}\left(q^{2}a_{\textbf{q}}^{2}-(\textbf{q}\textbf{a}_{\textbf{q}})^{2}\right)\right],
\end{equation}
where the last term is energy of the magnetic field
$\frac{1}{2\mu_{0}}\left(\texttt{curl}\textbf{A}\right)^{2}$.
Minimizing the free energy functional (\ref{3.8}) by the magnetic
field $\frac{\delta F}{\delta\textbf{a}_{\textbf{q}}}=0$ we obtain
a current in a form
\begin{equation}\label{3.9}
\textbf{j}_{\textbf{q}}=4eC|\Delta|\textbf{q}-8e^{2}C|\Delta|\textbf{a}_{\textbf{q}},
\end{equation}
where we can suppose $\Delta=\Delta_{\textbf{q}=0}=\frac{A}{B}$ in
the local (London) limit. In a case of a singly connected
superconductor using gauge transformations
$\textbf{a}_{\textbf{q}}\rightarrow
\textbf{a}_{\textbf{q}}+\textbf{q}/2e$ Eq.(\ref{3.9}) can be
reduced to the equation
$\textbf{j}_{\textbf{q}}=-8e^{2}C|\Delta|\textbf{a}_{\textbf{q}}$
describing Meissner's effect. The Londons' equation can be written
in a form
$\textbf{j}_{\textbf{q}}=-\frac{e^{2}}{m}2n_{s}\textbf{a}_{\textbf{q}}$,
where $n_{s}$ is a superfluid density (the density of correlated
pairs). Then
\begin{equation}\label{3.9a}
    n_{s}=\frac{\nu_{F}\hbar\omega
    k_{F}^{2}}{36\left(k_{B}T\right)^{3}m}\upsilon|\Delta|.
\end{equation}
Thus the superfluid density is $n_{s}\propto\upsilon|\Delta|$,
unlike BCS theory where $n_{s}\propto|\Delta|^{2}$. At the same
time, as in BCS theory, the response of a superconductor to
electromagnetic field is determined by the gap $\Delta$ which has
long-range coherence. If $\Delta=0$ then the pseudogap
$\upsilon/2$ cannot provide such electromagnetic response. General
expression for the superfluid density is given in Appendix
\ref{density}. If we consider a superconductor with an inner
cavity, then along a closed path lying within the superconductor
around the cavity at a distance from the cavity's surface larger
than magnetic penetration depth we have
$\textbf{j}_{\textbf{q}}=0\Rightarrow\textbf{q}=2e\textbf{a}_{\textbf{q}}
\Rightarrow\hbar\nabla\varphi=2e\textbf{A}$. Integrating along the
path we obtain magnetic flux quantization $\Phi=n\Phi_{0}$, where
$\Phi_{0}=\frac{\pi\hbar}{e}$ is a magnetic flux quantum. Thus the
free energy functional (\ref{3.8}) describes basic properties of
superconductivity.

\begin{figure}[h]
\includegraphics[width=7.5cm]{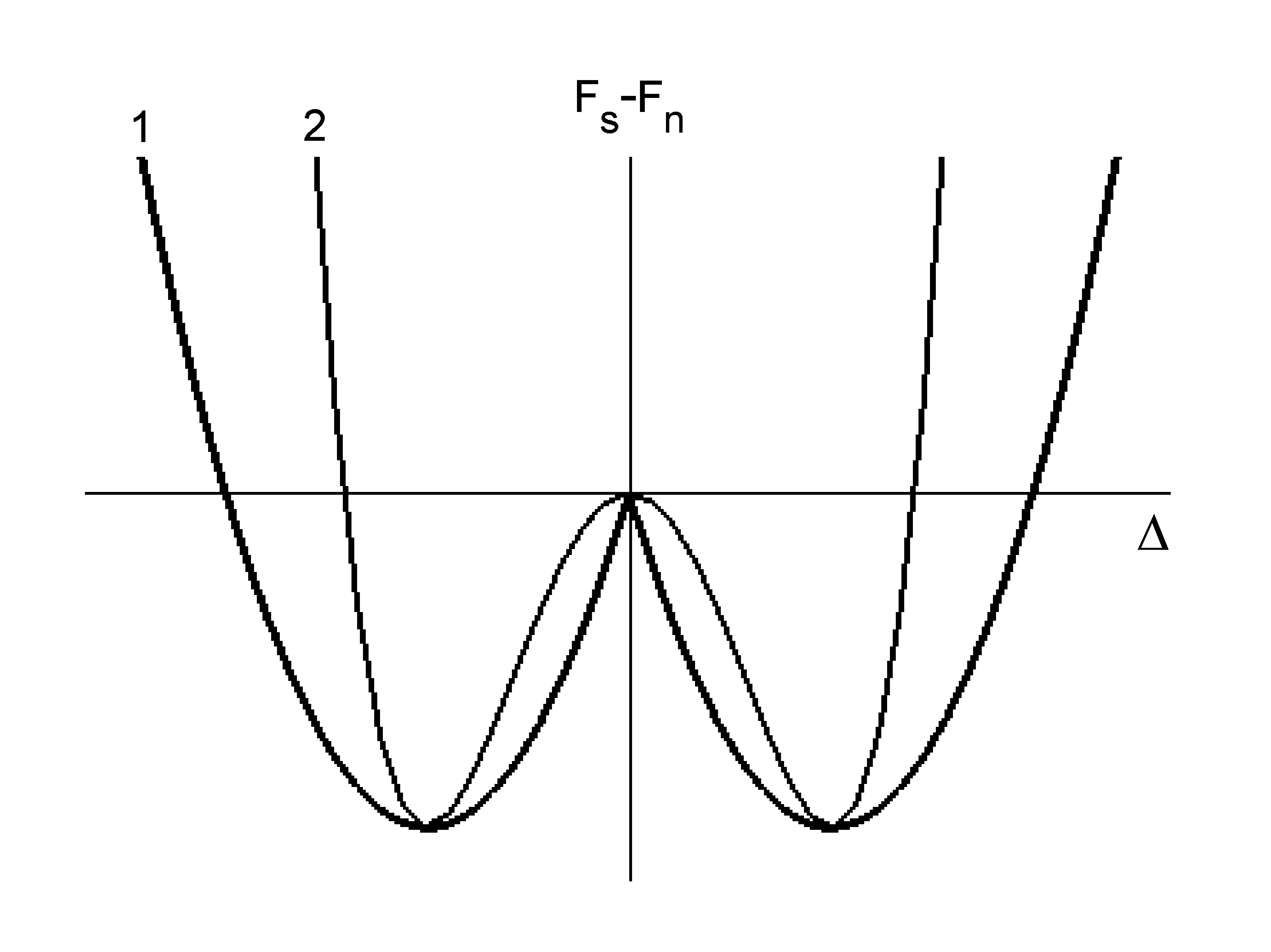}
\caption{the schematic illustration of the free energy (\ref{3.4})
at $\textbf{q}=0$ - line (1) and of the effective Ginzburg-Landau
free energy (\ref{3.10}) at
$\textbf{q}=0,\textbf{a}_{\textbf{q}}=0$ - line (2) which have the
same symmetry, the same extremes and the same values in these
extremes.} \label{Fig10}
\end{figure}

Using the transformations
$\Delta(\textbf{r})=\sum_{\textbf{q}}\Delta_{\textbf{q}}e^{i\textbf{qr}/\hbar},
\textbf{A}(\textbf{r})=\sum_{\textbf{q}}\textbf{a}_{\textbf{q}}e^{i\textbf{qr}/\hbar}$
and the inverse transformations
$\Delta_{\textbf{q}}=\frac{1}{V}\int\Delta(\textbf{r})e^{-i\textbf{qr}/\hbar}d\textbf{r},
\textbf{a}_{\textbf{q}}=\frac{1}{V}\int\textbf{A}(\textbf{r})e^{-i\textbf{qr}/\hbar}d\textbf{r}$
we can write the functional Eq.(\ref{3.8}) in real space, however
the functional will have a complicated and inconvenient form due
to terms $|\Delta_{\textbf{q}}|$ and $q^{2}|\Delta_{\textbf{q}}|$.
Nevertheless we have seen above that the functional (\ref{3.8})
describes basic properties of superconductivity, hence the
functional can be replaced by an effective GL functional, which
has the same symmetry, the same extremes and the same values in
these extremes - Fig.\ref{Fig10}. The effective GL functional has
a form
\begin{equation}\label{3.10}
F_{s}-F_{n}=V\sum_{\textbf{q}}\left[-B|\Delta_{\textbf{q}}|^{2}+\frac{B^{3}}{2A^{2}}|\Delta_{\textbf{q}}|^{4}
+\frac{BC}{A}\left(\textbf{q}-2e\textbf{a}_{\textbf{q}}\right)^{2}|\Delta_{\textbf{q}}|^{2}
+\frac{1}{2\mu_{0}\hbar^{2}}\left(q^{2}a_{\textbf{q}}^{2}-(\textbf{q}\textbf{a}_{\textbf{q}})^{2}\right)\right].
\end{equation}
The extremum $\Delta_{min}$, the value of the free energy
$(F_{s}-F_{n})_{min}$ and the critical momentum $q_{c}$ (if there
are only one $\textbf{q}$ in the sum and $\textbf{a}=0$) are
\begin{equation}\label{3.11}
|\Delta_{q}|_{min}=\frac{A}{B}\sqrt{1-\frac{C}{A}q^{2}},\quad
(F_{s}-F_{n})_{min}=-\frac{A^{2}}{2B}\left(1-\frac{C}{A}q^{2}\right)^{2}\Longrightarrow
q_{c}^{2}=\frac{A}{C}.
\end{equation}
Comparing Eq.(\ref{3.11}) with Eqs.(\ref{3.6},\ref{3.7}) we can
see coincidence of extremes $\Delta_{min}$ at $q=0$, coincidence
of the values $(F_{s}-F_{n})_{min}$ and coincidence of the
critical momentums $q_{c}$. Some different dependence
$\Delta_{min}(q)$ has no basic value. The current are
\begin{equation}\label{3.12}
\nonumber\textbf{j}_{\textbf{q}}=4e\frac{BC}{A}|\Delta|^{2}\textbf{q}-8e^{2}\frac{BC}{A}|\Delta|^{2}\textbf{a}_{\textbf{q}}
=4e\frac{AC}{B}\textbf{q}-8e^{2}\frac{AC}{B}\textbf{a}_{\textbf{q}},
\end{equation}
that coincides with the current (\ref{3.9}) in the local limit.
This means that the magnetic responses will be identical.

In real space the functional (\ref{3.10}) takes a form
\begin{equation}\label{3.13}
F_{s}-F_{n}=\int\left[-B|\Delta(\textbf{r})|^{2}+\frac{B^{3}}{2A^{2}}|\Delta(\textbf{r})|^{4}
+\frac{BC}{A}|\left(-i\hbar\nabla-2e\textbf{A}\right)\Delta(\textbf{r})|^{2}
+\frac{(\texttt{curl}\textbf{A})^{2}}{2\mu_{0}}\right]d\textbf{r},
\end{equation}
For convenience we can use a dimensionless variable
$\varphi(\textbf{r})\equiv\frac{\Delta(\textbf{r})}{\Delta_{0}}=\frac{\Delta(r)}{A/B}$,
then
\begin{equation}\label{3.14}
F_{s}-F_{n}=\int\left[\mu_{0}H_{cm}^{2}\left(-|\varphi(\textbf{r})|^{2}+\frac{1}{2}|\varphi(\textbf{r})|^{4}
+\xi^{2}\left|\left(\nabla-i\frac{2\pi}{\Phi_{0}}\textbf{A}\right)\varphi(\textbf{r})\right|^{2}\right)
+\frac{(\texttt{curl}\textbf{A})^{2}}{2\mu_{0}}\right]d\textbf{r},
\end{equation}
where
\begin{equation}\label{3.15}
   \mu_{0}H_{cm}^{2}=\frac{A^{2}}{B}\propto\frac{\upsilon^{2}}{T^{3}}
\end{equation}
is thermodynamical magnetic field $H_{cm}$,
\begin{equation}\label{3.16}
   \xi^{2}=\hbar^{2}\frac{C}{A}\propto\frac{1}{T}
\end{equation}
is a coherence length, which is determined with properties of a
superconductor only (at $T\rightarrow\infty$ it does not depend on
$\upsilon$). The coherence length in BCS (and in GL) theory
depends on temperature as $\xi=\xi_{0}/\sqrt{1-T/T_{c}}$. That is
at $T<T_{c}$ it increases as temperature rises, at $T=T_{c}$ it
diverges. At $T>T_{c}$ the coherence length has physical sense of
a correlation radius of fluctuations
$\xi\propto(T/T_{c}-1)^{-1/2}$ which decreases as $1/\sqrt{T}$ at
large $T$. The superconducting phase at $T>T_{c}$ arises
fluctuationally in a form of bubble of size $\xi$ within a normal
conductor, that is accompanied by an increasing of free energy
(the first term in Eq.(\ref{2.8})). The external pair potential
$\upsilon>0$ withholds the fluctuationally arisen superconducting
bubbles with such $\Delta$ that the increasing in the free energy
at formation of the babble is compensated by the potential
$\upsilon$ (the second term in Eq.(\ref{2.8})). This continues
until the superconducting phase does not fill the entire volume of
the metal. Thus the coherence length (\ref{3.16}) corresponds to
the correlation radius of fluctuations in GL theory. It should be
noted, that due to the transition to superconducting state in a
region of size $\xi(T)$ a change of the free energy is
\begin{equation}\label{3.16a}
    F_{s}-F_{n}\propto
    H_{cm}^{2}\xi^{3}(T)\propto\frac{\upsilon^{2}}{T^{9/2}}\ll
    k_{B}T,
\end{equation}
that is much less a thermal energy $k_{B}T$ at
$T\rightarrow\infty$. In GL theory the inequality (\ref{3.16a})
means strong fluctuations. In a case $\upsilon<0$ the situation is
opposite: the normal phase arises fluctuationally in the
superconductor, the external pair potential withholds the
fluctuationally arisen normal bubbles until the normal phase does
not fill the entire volume of the superconductor, thus
superconducting phase is suppressed - Fig.(\ref{Fig6}). This is
physical mechanism of the renormalization of the order parameter
(\ref{1.8},\ref{2.3}) by the external pair potential. Small
coherence length and strong fluctuations in this model shows
similarity with superfluid He II \cite{till}. Apparently due to
the strong fluctuations such regime of superconductivity is
possible in 3D systems only. Moreover Josephson effect cannot be
observed in such regime. In the same time we should notice that
the fluctuations at $T\rightarrow\infty, \upsilon>0$ do not
interact between themselves, since the free energy (\ref{3.8})
does not contain terms $\propto|\Delta_{\textbf{q}}|^{4}$, unlike
GL functional where the interaction of order parameter's
fluctuations plays principal role near the critical temperature:
$1-T/T_{c}<Gi$ \cite{lar}. Thus fluctuations in our model are
fundamentally different from fluctuations within the critical
domain in the standard theory.

It should be noted that phonons with lower energies than
temperature of the electron gas $\hbar\omega(q)\ll k_{B}T$ are
perceived by the electrons as static impurities \cite{ginz}. As
well known the static impurities do not influence on value of
$T_{c}$ in $s$-wave superconductors (the Anderson's theorem).
However for $d$-wave pairing the nonmagnetic impurities destroy
superconductivity like magnetic impurities, hence suppression of
superconductivity by thermal phonons is observed in such systems
and a change to the first order transition occurs
\cite{gab1,gab2}. Thus above-described regime of superconductivity
is possible in $s$-wave superconductors only.

Varying the functional (\ref{3.14}) by $\varphi^{+}$ and
$\textbf{A}$ we obtain GL equations
\begin{equation}\label{3.17}
    \left\{\begin{array}{c}
      -\xi^{2}\left(\nabla-i\frac{2\pi}{\Phi_{0}}\textbf{A}\right)^{2}\varphi-\varphi+\varphi|\varphi|^{2}=0 \\
       \\
      \texttt{curl}\texttt{curl}\textbf{A}
      =-i\frac{\Phi_{0}}{4\pi\lambda^{2}}\left(\varphi^{+}\nabla\varphi-\varphi\nabla\varphi^{+}\right)
      -\frac{|\varphi|^{2}}{\lambda^{2}}\textbf{A} \\
    \end{array}\right\},
\end{equation}
where a magnetic field penetration depth is
\begin{equation}\label{3.18}
    \frac{1}{\lambda^{2}}=\frac{8\pi^{2}\mu_{0}^{2}H_{cm}^{2}\xi^{2}}{\Phi_{0}^{2}}\Rightarrow\lambda\propto\frac{T^{2}}{\upsilon}.
\end{equation}
The penetration depth increases with temperature, however, unlike
the ordinary GL theory, it is finite quantity at all temperatures.
Then GL parameter is
\begin{equation}\label{3.19}
\kappa=\frac{\lambda}{\xi}\propto\frac{T^{5/2}}{\upsilon}.
\end{equation}
The parameter increases with temperature unlike the ordinary GL
theory, where the parameter is constant. This means that at large
temperature all superconductors under the external pair potential
become type II superconductors. Knowing the thermodynamical
magnetic field and the magnetic penetration depth we obtain a
pair-breaking current as \cite{schmidt}:
\begin{equation}\label{3.19a}
    j_{c}=\frac{2\sqrt{2}}{3\sqrt{3}}\frac{H_{cm}}{\lambda}\propto\frac{\upsilon^{2}}{T^{7/2}},
\end{equation}
Besides the critical temperature an important characteristic of a
superconductor are the first $H_{c1}$ and the second $H_{c2}$
critical fields. The second critical field is
\begin{equation}\label{3.20}
    H_{c2}=\frac{\Phi_{0}}{2\pi\mu_{0}\xi^{2}}=\sqrt{2}\kappa H_{cm}\propto T.
\end{equation}
$H_{c2}$ is proportional to temperature that differs radically
from GL theory, where $H_{c2}\propto 1-T/T_{c}$. In GL theory the
transition to normal state in magnetic field occurs when average
distance between vortexes becomes the order of the coherence
length $\xi$. However in our case the length decreases with
temperature $\xi\propto 1/\sqrt{T}$. This decrease compensates the
rapprochement of the vortexes. The first critical field is
\begin{equation}\label{3.21}
    H_{c1}=\frac{H_{c2}}{2\kappa^{2}}\ln\kappa\propto\frac{\upsilon^{2}}{T^{4}}.
\end{equation}
We can see the relatively strong fall with temperature, that
indicates the above-mentioned tendency of transition to type II
superconductivity at large temperature. The field of melting of
the vortex lattice \cite{schmidt} is
\begin{equation}\label{3.22}
    H_{m}\propto\frac{\Phi_{0}}{(k_{B}T)^{2}}\left(\frac{\Phi_{0}}{\lambda}\right)^{4}\propto\frac{\upsilon^{4}}{T^{10}}.
\end{equation}
Obviously at $T\rightarrow\infty$ the field decreases with
temperature faster than the first critical field $H_{m}\ll
H_{c1}\ll H_{c2}$ as temperature increases. Thus in this regime
the vortex lattice is not formed, that is result of
above-mentioned strong fluctuations.

\section{Summary}\label{result}

In this work we have considered a hypothetical substance, where an
interaction energy between (within) structural elements of
condensed matter (molecules, nanoparticles, clusters, layers,
wires etc.) depends on state of Cooper pairs: an additional work
$\upsilon$ must be made against this interaction to break a pair.
In this case the invariant under $U(1)$ transformation source of
the pairs $\widehat{H}_{\upsilon}$ has to be added to BCS
Hamiltonian. We have obtained a free energy for this model, which
has two minimums. One of them corresponds to injection of the
pairs into the system from an external source. Even if the
electron-electron interaction is absent the energy gap is nonzero
$|\Delta|=\upsilon/2$ (if $\upsilon>0$) that corresponds to
results obtained in works \cite{loz,cap}. In another minimum
(\ref{1.8},\ref{2.3}) only the electron-electron coupling is the
cause of superconductivity, but not the source
$\widehat{H}_{\upsilon}$, however \emph{the source essentially
renormalizes the order parameter}: in presence of the potential
$\upsilon\neq 0$ the self-consistency condition has a form
$\Delta=I(\Delta,\upsilon)$. In this case we call the source as
the \textit{external pair potential}, since the potential is
imposed on the electron subsystem by the structural elements of
matter, unlike the pair potential $\Delta$, which is result of
electron-electron interaction and determined with the
self-consistency equation. If $\upsilon=0$ we have an usual
self-consistency equation in BCS theory. If $\upsilon<0$ then the
pairing of electrons increases energy of the molecular structure,
that suppresses superconductivity. The phase transition
conductor-superconductor becomes the first order phase transition
with lower critical temperature than at $\upsilon=0$ and a region
of metastable states occurs. If $\upsilon>0$ then the pairing
lowers the energy of the molecular structure, that supports
superconductivity. In this case at large temperatures $T\gg T_{c}$
the energy gap tends to zero asymptotically as $1/T$. Thus,
formally, the critical temperature is equal to infinity, however
the energy gap remains finite quantity. Hence the ratio between
the gap and the critical temperature is $\Delta/T_{c}=0$ instead
of finite values $\sim 1$ for all known materials. Possible
realization of this model has been proposed in \cite{grig2}. It
should be noted that the normal state $\Delta=0$ in this model
(when $\upsilon>0$) can be considered as state with a pseudogap:
since the quasiparticle spectrum $E_{0}$ has a gap $\upsilon/2$,
then charge is carried by the pairs of electrons, but this state
is not superconducting because the ordering $\left\langle
aa\right\rangle,\left\langle a^{+}a^{+}\right\rangle$, i.e.
long-range coherence, is absent. The uncorrelated pairs are pairs
in momentum space (unlike bipolarons which are pairs in real
space) which exist in noninteracting Fermi system due to the
external pair potential. Switching-on of the electron-electron
interaction stipulates the phase coherence. The superfluid density
$n_{s}$ is proportional to $\upsilon|\Delta|$, unlike BCS theory
where $n_{s}\propto|\Delta|^{2}$. At the same time, as in BCS
theory, the response of a superconductor to electromagnetic field
is determined by the gap $\Delta$ which has long-range coherence,
the pseudogap $\upsilon/2$ does not give any contribution to such
electromagnetic response.

For the case $\upsilon>0$ and $T\rightarrow\infty$ the effective
GL free energy functional has been obtained. The functional
describes a superconductor without critical temperature: the order
parameter depends on temperature as $1/T$. The thermodynamical
magnetic field $H_{cm}$, the first critical field $H_{c1}$ and the
pair-breaking current $j_{c}$ tend to zero asymptotically as
temperature increases and they are functions of the external pair
potential. The magnetic penetration depth $\lambda$ increases with
temperature as $T^{2}$ and it is a finite quantity at all
temperatures. The coherence length $\xi$ decreases at large
temperature as $1/\sqrt{T}$. The length corresponds to a size of
fluctuationally arising bubbles of superconducting phase in a
normal conductor (the correlation radius of fluctuations). At
$T>T_{c}$ the superconducting phase arises fluctuationally in a
normal conductor. The external pair potential $\upsilon>0$
withholds the fluctuationally arisen superconducting bubbles. This
continues until the superconducting phase does not fill the entire
volume of the metal. In this case a change of the free energy due
to transition to superconducting state in the region of size
$\xi(T)$ is much less a thermal energy $k_{B}T$, that means strong
fluctuations. In a case $\upsilon<0$ the external pair potential
withholds the fluctuationally arisen normal bubbles in a
superconductor until the normal phase does not fill the entire
volume of the superconductor, thus superconducting phase is
suppressed. This is physical mechanism of above-mentioned
renormalization of the order parameter by the external pair
potential. Small coherence length and strong fluctuations in this
model shows similarity with superfluid He II. Apparently due to
the strong fluctuations such superconductivity is possible in 3D
systems only. In the same time the fluctuations at
$T\rightarrow\infty, \upsilon>0$ do not interact between
themselves. Thus fluctuations in our model are fundamentally
different from fluctuations within the critical domain in the
standard GL theory. Moreover such superconductivity is possible in
$s$-wave superconductors only due to large numbers of thermal
phonons which are perceived by the electrons as static impurities.

Due to the coherence length is a decreasing function of
temperature $\xi\propto 1/\sqrt{T}$ the GL parameter
$\kappa=\lambda/\xi$ is an increasing function of
temperature $\propto T^{5/2}$ unlike the ordinary GL theory, where
the parameter is constant. Hence at large temperature all
superconductors under the external pair potential become type II
superconductors with the second critical field $H_{c2}$
proportional to temperature. In GL theory the
transition to normal state in magnetic field occurs when average
distance between vortexes becomes order of the coherence length
$\xi$. However in our model the length decreases with temperature,
that compensates the rapprochement of the vortexes. This result
differs from results of a work \cite{grig1}, where the second
critical field is infinity like the critical temperature. The free
energy functional (\ref{3.14}) has been obtained in
high-temperature limit from the general expression (\ref{2.5}). In
contrast, in \cite{grig1} the free energy functional was obtained
by modifying of GL expansion. Another feature of this model is
that the field of melting of the vortex lattice is
$H_{m}\ll H_{c1}$  at $T\rightarrow\infty$. Thus
in this regime the vortex lattice is not formed, that is a result
of the strong fluctuations in such a system.

Thus we propose a fundamentally different approach to the problem
of increasing of critical temperature. This approach is not
associated with enlargement of the coupling constant or with
change of the frequency, but it allows to reformulate the problem
in the sense that we change the ratio between the energy gap and
the critical temperature as $\Delta/T_{c}\rightarrow 0$ at finite
value of the gap. Since in this model the second critical field is
an increasing function of temperature this gives opportunity to
have a superconducting state at very large magnetic fields.

\appendix
\section{Superfluid density}\label{density}

Hamiltonian of particles with spectrum $\varepsilon(\textbf{k})$
in the external magnetic field is
\begin{equation}\label{A1}
    \widehat{H}=\int\psi^{+}(\textbf{r})\varepsilon\left(\widehat{\textbf{q}}-e\textbf{A}\right)\psi(\textbf{r})d\textbf{r},
\end{equation}
where $\widehat{\textbf{q}}=-i\hbar\nabla_{\textbf{r}}$. Varying
the Hamiltonian with respect to \textbf{A} we obtain:
\begin{equation}\label{A2}
    \widehat{\textbf{j}}=e\psi^{+}(\textbf{r})\nabla_{\textbf{q}}\varepsilon\left(\textbf{q}-e\textbf{a}\right)\psi(\textbf{r}),
\end{equation}
where
$\textbf{a}_{\textbf{q}}=\frac{1}{V}\int\textbf{A}(\textbf{r})e^{-i\textbf{qr}/\hbar}d\textbf{r}$.
Let us leave a linear relative to \textbf{a} term:
\begin{equation}\label{A3}
\widehat{\textbf{j}}=e\psi^{+}(\textbf{r})\nabla_{\textbf{q}}\varepsilon(\textbf{q})\psi(\textbf{r})
-e^{2}\psi^{+}(\textbf{r})\nabla_{\textbf{q}}\left[(\textbf{a}\nabla_{\textbf{q}})\varepsilon(\textbf{q})\right]\psi(\textbf{r}).
\end{equation}
Here the first term is a paramagnetic part and the second term is
diamagnetic part of the current. Then connection between current
and magnetic field takes a form:
\begin{equation}\label{A4}
    \textbf{j}=Q(\textbf{q})\textbf{a}
\end{equation}
where $Q(\textbf{q})=Q(0)+c\textbf{q}^{2}+\ldots$. For normal
metal $Q(0)=0$, that is the diamagnetic part is exactly
compensated by the paramagnetic part (Ward's identity). In a
superconductor this identity is violated:
\begin{equation}\label{A5}
    Q(0)=\frac{e^{2}}{m}2n_{s},
\end{equation}
where $n_{s}$ is the superfluid density - number of correlated
pairs. Following \cite{levit,sad} we write the superfluid density
in a form:
\begin{equation}\label{A6}
    2n_{s}=n_{0}k_{B}T\sum_{n}\int_{-\infty}^{+\infty}\left[\frac{\varepsilon^{2}-\omega_{n}^{2}}{\left(\varepsilon^{2}+\omega_{n}^{2}\right)^{2}}-
    \frac{\varepsilon^{2}-\omega_{n}^{2}+|\Delta|^{2}}{\left(\varepsilon^{2}+\omega_{n}^{2}+|\Delta|^{2}\right)^{2}}\right]d\varepsilon
    =n_{0}k_{B}T\sum_{n}\frac{\pi|\Delta|^{2}}{\left(\omega_{n}^{2}+|\Delta|^{2}\right)^{3/2}},
\end{equation}
where
$n_{0}=\frac{p_{F}^{3}}{3\pi^{2}\hbar^{3}}=\frac{1}{3\pi^{2}}\left(\frac{2m\varepsilon_{F}}{\hbar^{2}}\right)^{3/2}$
is a total density of conduction electron, $\omega_{n}=\pi
k_{B}T(2n+1)$. The first term in the square brackets corresponds
to the normal state of metal only, it compensates contribution of
normal electrons in the second term. For case $T=0$ the superfluid
density is equal to half of the total electron density
$n_{s}=n_{0}/2$ as it must be in a translationally invariant
system. For case $T\rightarrow T_{c}$, where $\Delta\rightarrow
0$, we have
\begin{equation}\label{A7}
    n_{s}=n_{0}\frac{7|\Delta|^{2}}{8\pi^{2}(k_{B}T_{c})^{2}}=n_{0}\left(1-\frac{T}{T_{c}}\right).
\end{equation}
Thus at $T\rightarrow T_{c}$ the superfluid density decreases to
zero.

For the BCS theory with the external pair potential we must take
into account that the spectrum of quasiparticles is
$E^{2}=\varepsilon^{2}+|\Delta|^{2}\left(1+\frac{\upsilon}{2|\Delta|}\right)^{2}$,
hence the spectrum in the normal state is
$E_{0}^{2}=\varepsilon^{2}+\left(\frac{\upsilon}{2}\right)^{2}$
(state with "pseudogap"). Then the superfluid density is
\begin{eqnarray}\label{A8}
    2n_{s}&=&n_{0}k_{B}T\sum_{n}\int_{-\infty}^{+\infty}\left[\frac{E_{0}^{2}-\omega_{n}^{2}}{\left(E_{0}^{2}+\omega_{n}^{2}\right)^{2}}-
    \frac{E^{2}-\omega_{n}^{2}}{\left(E^{2}+\omega_{n}^{2}\right)^{2}}\right]d\varepsilon\nonumber\\
    \nonumber\\
    &=&n_{0}k_{B}T\left[\sum_{n}\frac{\pi\left(|\Delta|^{2}+|\Delta|\upsilon+\left(\frac{\upsilon}{2}\right)^{2}\right)}
    {\left(\omega_{n}^{2}+\left(|\Delta|^{2}+|\Delta|\upsilon+\left(\frac{\upsilon}{2}\right)^{2}\right)\right)^{3/2}}
    -\sum_{n}\frac{\pi\left(\frac{\upsilon}{2}\right)^{2}}{\left(\omega_{n}^{2}+\left(\frac{\upsilon}{2}\right)^{2}\right)^{3/2}}\right],
\end{eqnarray}
Analogously to the previous case we have $n_{s}=n_{0}/2$ for the
case $T=0$. For the case $T\rightarrow \infty$, where
$\Delta\rightarrow 0$, we have
\begin{equation}\label{A9}
    n_{s}=n_{0}\frac{7\upsilon|\Delta|}{8\pi^{2}(k_{B}T)^{2}},
\end{equation}
that is in qualitative agreement with (\ref{3.9a}): the superfluid
density is proportional to $\upsilon|\Delta|$, unlike BCS theory
where $n_{s}\propto|\Delta|^{2}$. At the same time, as in BCS
theory, the response of a superconductor to electromagnetic field
is determined by the gap $\Delta$ which has long-range coherence.
If $\Delta=0$ then the pseudogap $\upsilon/2$ cannot provide such
electromagnetic response.

In Section \ref{zero} we could see that the normal state is a
state with a pseudogap: if the ordering $\Delta\propto\left\langle
aa\right\rangle,\Delta^{+}\propto\left\langle
a^{+}a^{+}\right\rangle$, i.e. long-range coherence, is absent,
then the quasiparticle spectrum has a gap $\upsilon/2$. This means
that charge is still carried by the pairs of electrons but the
pairs are not correlated. By analogy with Eqs.(\ref{A6},\ref{A8})
the total density of pairs $n_{p}$ (correlated + uncorrelated
pairs) can be found as
\begin{eqnarray}\label{A10}
    2n_{p}&=&n_{0}k_{B}T\sum_{n}\int_{-\infty}^{+\infty}\left[\frac{\varepsilon^{2}-\omega_{n}^{2}}{\left(\varepsilon^{2}+\omega_{n}^{2}\right)^{2}}-
    \frac{E^{2}-\omega_{n}^{2}}{\left(E^{2}+\omega_{n}^{2}\right)^{2}}\right]d\varepsilon,
\end{eqnarray}
As in the previous cases we have $n_{p}=n_{0}/2$ for case $T=0$.
For case $T\rightarrow \infty$, where $\Delta\rightarrow 0$, we
obtain
\begin{eqnarray}\label{A11}
    n_{p}=n_{0}\frac{7|\Delta|^{2}\left(1+\frac{\upsilon}{2|\Delta|}\right)^{2}}{8\pi^{2}(k_{B}T)^{2}}
    \approx n_{0}\frac{7\upsilon^{2}}{32\pi^{2}(k_{B}T)^{2}}.
\end{eqnarray}
Thus, the total density of pairs is proportional to $\upsilon^{2}$
(it is nonzero even when $\Delta=0$) and it tends to zero with
increasing temperature slower than the superfluid density.
However, the presence of the pseudogap raises the question about
the fermionic or bosonic nature of Cooper pairs. Following
\cite{dzum} we suppose that if the size of a Cooper pair $a_{p}$
is much larger than the mean distance $R_{p}$ between the Cooper
pairs then the bosonization of such Cooper pairs cannot be
realized due to their strong overlapping. Thus for fermionic
nature of the pair it should be $a_{p}/R_{p}>1$. Using the
uncertainty principle, the size of the polaronic Cooper pairs is
defined as
$a_{p}(T)=\left(\hbar/2|\widetilde{\Delta}|\right)\sqrt{\varepsilon_{F}/2m}$,
where $|\widetilde{\Delta}|$ is the energy gap (in our case
$|\Delta|+\upsilon/2$). The size is compared with $R_{p}=(3/4\pi
n_{p})^{1/3}$, where $n_{p}$ is the density of the pairs
(\ref{A10},\ref{A11}). Then for zero temperature we have
$a_{p}/R_{p}\sim\varepsilon_{F}/\upsilon\gg 1$. For the case
$T\rightarrow\infty$ we obtain the condition for bosonization of
Cooper pairs in a form
$k_{B}T>\varepsilon_{F}^{3/2}/\upsilon^{1/2}$. For evaluation let
us consider a reasonable value $\upsilon\sim 10\texttt{K}$ (usual
bound energy of the pairs) and $\varepsilon_{F}\sim
0.2\texttt{eV}$ (superconductors with low density of carriers, for
example alkali-doped fullerides), then we have
$T>10^{4}\texttt{K}$ that far exceeds the melting point of the
medium. In metals this temperature is even higher. Hence the
Cooper pairs have fermionic nature. Therefore the superconducting
properties (critical temperature $T_{c}$ and critical magnetic
fields $H_{c1}$, $H_{c2}$) can be described by the generalized BCS
theory.

\acknowledgments

The work is supported by the project 0113U001093 of the National
Academy of Sciences of Ukraine.



\end{document}